\documentclass{pasj01}
\draft 
\Received{$\langle$reception date$\rangle$}
\Accepted{$\langle$acception date$\rangle$}
\Published{$\langle$publication date$\rangle$}
\usepackage{lineno}
\begin{document}

\title{Effects of Depolarizing Intervening Galaxies on Background Radio Emission I. Global Disk Magnetic Field}

%%% begin:list of authors
% Do NOT capitalize all letters in "textsc".
\author{Rikuto \textsc{omae} \altaffilmark{1,2,*}}
\author{Takuya \textsc{akahori}\altaffilmark{3,4}}
\author{Mami \textsc{machida}\altaffilmark{5}}
\altaffiltext{1}{Department of Astronomical Science, School of Physical Sciences, The Graduate University for Advanced Studies (SOKENDAI), 2-21-1 Osawa, Mitaka, Tokyo 181-8588, Japan}
\altaffiltext{2}{Department of Physics, Faculty of Sciences, Kyushu University, 744 Motooka, Nishi-ku, Fukuoka 819-0395, Japan}
\altaffiltext{3}{Mizusawa VLBI Observatory, National Astronomical Observatory of Japan, 2-21-1, Osawa, Mitaka, Tokyo 181-8588, Japan}
\altaffiltext{4}{Operation Division, Square Kilometre Array Observatory, Lower Withington, Macclesfield, Cheshire SK11 9FT, UK}
\altaffiltext{5}{Division of Science, National Astronomical Observatory of Japan, 2-21-1 Osawa, Mitaka, Tokyo 181-0015, Japan}
\email{rikuto.omae@grad.nao.ac.jp}
%%% end:list of authors

\KeyWords{galaxies: general --- magnetic fields --- polarization}
\maketitle

\begin{abstract}
External galaxies often intervene in front of background radio sources such as quasars and radio galaxies. Linear polarization of the background emission is depolarized by Faraday rotation of inhomogeneous magnetized plasma of the intervening galaxies. Exploring the depolarizing intervening galaxies (DINGs) can be a powerful tool to investigate the cosmological evolution of the galactic magnetic field. In this paper, we investigate the effects of DINGs on background radio emission using theoretical DING models. We find that complex structures of galaxy result in complicated depolarization features and the Faraday dispersion functions (FDFs), but, for the features of depolarizations and FDFs, the global component of magnetic fields is important. We show the simplest results with ring magnetic field in the galactic disk. We find that the degree of depolarization significantly depends on the inclination angle and the impact parameter of the DING. We found that the larger the standard deviation, the more likely it is that depolarization will occur. The FDF represents the RM structure within the beam. The FDF exhibits multi-components due mainly to the RM structure within the beam and the fraction of the DING that covers the background emission (the filling factor). The peak Faraday depth of the FDF is different from the beam-averaged RM of the DING. The Monte-Carlo simulations indicate that DING's contribution to the standard deviation of observed RMs follows $\sigma_{\rm RM} \propto 1/{(1+z)^k}$ with $k \sim 2.7$ and exhibits a steeper redshift dependence than the wavelength squared. DINGs will have a significant impact on RM catalogs created by future survey projects such as the SKA and SKA Precursor/Pathfinder.

\end{abstract}
%\linenumbers
\section{Introduction}\label{intro}

Magnetic fields play a fundamental role in various astrophysical processes through magnetic tension, instability, turbulence, heat dissipation, and cosmic-ray acceleration. These processes impact on the galaxy formation and evolution (e.g., \cite{2009ASTRA...5...43B, 2013MNRAS.432..176P}). The galactic magnetic fields have been studied through observations of diffuse synchrotron emission and the Zeeman effect of line emission (see, \cite{2015A&ARv..24....4B} for a review). Those observations are powerful, but applicable for bright galaxies, remaining the bias of sample selection. The cosmological evolution of galactic magnetic field is largely unknown because these emissions from distant galaxies are too faint to observe.

One of the possible, less-biased ways to study galactic magnetic fields is to investigate the depolarizing intervening galaxies (hereafter, DINGs). There are a lot of external galaxies in the intergalactic space, so that the galaxies can intervene in the emission from background astronomical objects such as quasars and radio galaxies. Indeed, 40,429 out of 84,533 SDSS quasars are known to accompany Mg II absorber systems in front of the quasars (\cite{2013ApJ...770..130Z}). Such absorber systems are likely DINGs, whose magnetic fields can depolarize background polarized emission and can contribute to Faraday rotation measure (RM). This overlap happens randomly in space, reducing the selection bias. The correlation between polarization properties and the existence of Mg II absorber systems has been reported in the literature \citep{2008ApJ...676...70K, 2008Natur.454..302B, 2012ApJ...761..144B, 2014ApJ...795...63F, 2020ApJ...890..132M}. For example, \citet{2014ApJ...795...63F} reported that the subset sample showing flat radio spectra indicates significant evidence for a correlation between Mg II absorption and RM at 1.4 GHz, while that with steep spectra shows no such a correlation. Large surveys of extragalactic sources with the Square Kilometre Array (SKA) (e.g., \cite{2004NewAR..48.1003G}) and its precursors/pathfinders will dramatically increase the number of extragalactic polarized sources, allowing us to improve previous statistical works for DINGs. 

Much work has been devoted to understand DINGs using various simple models (e.g., \cite{1966MNRAS.133...67B}; \cite{1998MNRAS.299..189S}) and galaxy models (e.g., \cite{2009ApJ...693.1392S, 2012A&A...543A.127S}; \cite{2012ApJ...761..144B}). However, redshift dependence of depolarization caused by a DING is not systematically addressed, remaining statistical contribution of DINGs in RM catalogs unclear. In addition, while modern radio polarimetry adopts Faraday tomography to estimate RM (e.g., \cite{2017A&A...597A..98V, 2018Galax...6..137S}, \cite{2020MNRAS.495..143C}), DING's effects on Faraday dispersion function (FDF) is not paid much attention \citep{2016ApJ...829..133K}. Clarification of DING's effects is crucial for exploring galactic \citep{2008A&A...477..573S, 2012A&A...542A..93O, 2013ApJ...767..150A, 2014RAA....14..942X, 2021arXiv210201709H} and intergalactic \citep{2010ApJ...723..476A, 2011ApJ...738..134A, 2012arXiv1209.1438H, 2019ApJ...878...92V, 2020MNRAS.495.2607O} magnetic fields with future large surveys.

In this paper, we show depolarized spectra and FDFs caused by the DINGs with various physical parameters, and explore their statistical significance for large surveys of RM in future, using theoretical DING models. The paper is organized as follows. We briefly summarize the nature of depolarization in Section \ref{Breif_Depo} and describe our model and calculation in Section \ref{model_cal}. We find that, while multiple galactic structures result in complicated depolarization features, the essence is made by the global disk magnetic field which is the primary component of galactic RM. Based on this fact, we show the results for the simplest, ring magnetic field in the galactic disk to clarify how DING's depolarization takes place in Section \ref{Result}. Other galactic components such as the disk spiral, the halo, and turbulent magnetic fields are added in our discussion (Section \ref{Discuss}). We also discuss systematic errors on estimating DING's RM mean and dispersion by the FDF peak and the Burn's depolarization law, respectively. Finally, we discuss redshift dependence of the DING's RM dispersion and a future prospect to study the intergalactic magnetic field. We summarize this work in Section \ref{conc}.

\section{Brief Summary of Depolarization }\label{Breif_Depo}

Depolarization is a mechanism by which an observer receives a weaker polarized emission than that emitted at the origin. It leads to a decrease in the polarization fraction since this mechanism does not affect the total intensity. Because depolarization is common in many astronomical cases, the properties of depolarization can be a useful tool for revealing the structure of the line of sight and the source itself. Analytical works have demonstrated various possible cases of depolarization (e.g., \cite{1966MNRAS.133...67B}, \cite{1991MNRAS.250..726T}, \cite{1998MNRAS.299..189S}, \cite{2011MNRAS.418.2336A}, see also \cite{2018PASJ...70R...2A} for a review). Because diffuse polarized emission of a DING is expected to be several orders of magnitude weaker than those of quasars and radio galaxies, we can safely neglect the polarized emission of the DING. Thus, the DING plays a role of Faraday screen and the external Faraday dispersion depolarization or the beam depolarization emerges \citep{1998MNRAS.299..189S}.

Faraday screen consists of magnetized plasma that causes Faraday rotation. If there are only coherent magnetic fields and a constant RM structure in this Faraday screen, the emission of the background source experiences the same amount of Faraday rotation and the screen does not cause any depolarization. On the other hand, if the emission experiences different amounts of Faraday rotation at different positions, cancellation of the linear polarization can take place. This is the beam depolarization and is considered to be the primary mechanism of depolarization. Depolarization is thus more likely to occur when there is a RM gradient and/or there are turbulent magnetic fields within an observing radio beam (see, e.g., \cite{1998MNRAS.299..189S}). In this paper we focus on the former effect and the latter effect will be studied in the following paper II (Omae et al. in preparation).

In the case of the beam depolarization, the effect of depolarization is greater at longer wavelengths in general, because the effect of Faraday rotation is stronger at longer wavelengths. If the relationship between the total intensity $I(\nu) \propto \nu^{\alpha_I}$ and the linearly-polarized intensity $P(\nu) \propto \nu^{\alpha_P}$ holds, the most general expression for depolarization can be expressed as follows;
\begin{equation}
p = P/I \propto \nu^{\alpha_P-\alpha_I}=\nu^\beta,
\end{equation}
where $p$ is the polarization fraction, $\nu$ is the observed frequency, $\alpha_I$ is the total intensity spectral index, $\alpha_P$ is the polarization spectral index, and $\beta = \alpha_P-\alpha_I$. In reality, the emission of the background source is considered to be non-uniform and depolarized by itself. We will address this effects in future work.

The polarization that passes through regions of large RM tends to be more depolarized than that passes through regions of small RM, so that the polarization fraction is biased and the observed RM tends to be smaller than the average of intrinsic values within the beam. Furthermore, an observed RM is always smaller than the intrinsic value by $1/(1 + z)^2$, where $z$ is the redshift at which Faraday rotation happens because the rotation takes place at a higher rest-frame frequency. 

\section{Model and Calculation}\label{model_cal}

In this section, we describe our model and calculation. When we refer the angular size, we adopt a $\Lambda \rm{CDM}$ cosmology with ${\rm \Omega_{m0}}= 0.27$ and ${\rm \Omega_{\Lambda 0}}= 0.73$, and ${\rm H_0} = 70\, \mathrm{km\, s^{-1}Mpc^{-1}}$ in this paper. One arcsecond ($1^{\prime\prime}$) corresponds to 1.8, 6.2, and 8.2 kpc for redshifts of $z$ = 0.1, 0.5, and 1.0, respectively.

\subsection{Numerical Models}

We consider observing a radio lobe as a background polarized source behind a DING. We only consider RM of the DING and ignore any other RMs such as RMs of a background source and the Milky Way Galaxy; those are beyond the scope of this paper. Although a background radio source likely has inhomogeneous morphology, we assume that the source has uniform emission and a circular shape with 1$^{\prime\prime}$ in diameter to focus on the effect of the DING.

We define the Stokes parameters, $I(\nu)$, $Q(\nu)$, $U(\nu)$, $V(\nu)$, and the polarized intensity, $P(\nu)=Q(\nu)+j U(\nu)$, of a background source as
\begin{equation}
I(\nu) = I_0\nu^{\alpha_I},
\end{equation}
\begin{equation}
 P(\nu) = P_0\nu^{\alpha_P}=p (\nu)I(\nu)e^{2j\chi_0},
\end{equation}
where $\chi_0$ is the intrinsic polarization angle, $j$ is the imaginary unit, and we consider no circular polarization, i.e. $V(\nu)=0$. For the intrinsic emission of background source, we adopt the simplest parameters; $\alpha_I=\alpha_P=-1$ and $p_0=P_0/I_0=1$. The choice of the parameters is useful to visualize depolarization effects of our interest in this paper, and does not affect our conclusion of depolarization effects caused by a DING.

DINGs are external galaxies that are dark in any observing frequency but are depolarizing background polarization. We consider a low surface brightness, disk galaxy as a DING since the field sky dominates a large survey volume. Magnetic fields in disk galaxies have been investigated in the literature (e.g., \cite{1986ARA&A..24..459S, 2001SSRv...99..243B, 2008A&A...477..573S, 2009ASTRA...5...43B}). They consist of global, regular components and local, turbulent components, where we only consider the former component as stated in Section \ref{intro}.

We adopt the RING model as a global regular magnetic field of a DING {(e.g., \cite{1990IAUS..140..227S})}. The ring-shape magnetic field can be written in the cylindrical coordinate system $(R, \phi, z^{\prime})$ as
\begin{eqnarray}
  \left\{
    \begin{array}{l}
      B_R = 0, \\
      B_\phi = B_0\ (0~{\rm kpc} \leq R \leq10~{\rm kpc}), \\
      B_{z^{\prime}} = 0,
    \end{array}
  \right.
\end{eqnarray}
where we adopt the disk scale height of 1 kpc and the strength of $B_0$  $=5~\mathrm{\mu G}$;
\begin{eqnarray}
  B_0 = \left\{
    \begin{array}{l}
      5\ (-1~{\rm kpc} \leq z^{\prime} \leq 1~{\rm kpc}), \\
      0\ (z^{\prime} < -1~{\rm kpc}, 1~{\rm kpc}< z^{\prime}). \\
    \end{array}
  \right.
\end{eqnarray}

The global electron distribution is assumed to be uniform with the density of $n_{\rm e}=0.01\ {\rm cm^{-3}}$. This assumption is reasonable while we consider the disk component up to the scale height of the disk. We do not consider synchrotron emission from cosmic-rays in the DING because it is negligibly weak by the definition of the DING. For the study of FDF of diffuse polarized emission from external galaxies, see e.g., \citet{2020ApJ...899..122E} and references therein.

\begin{figure}
  \begin{center}
  	\includegraphics[width=1\linewidth,bb=0 0 849 432]{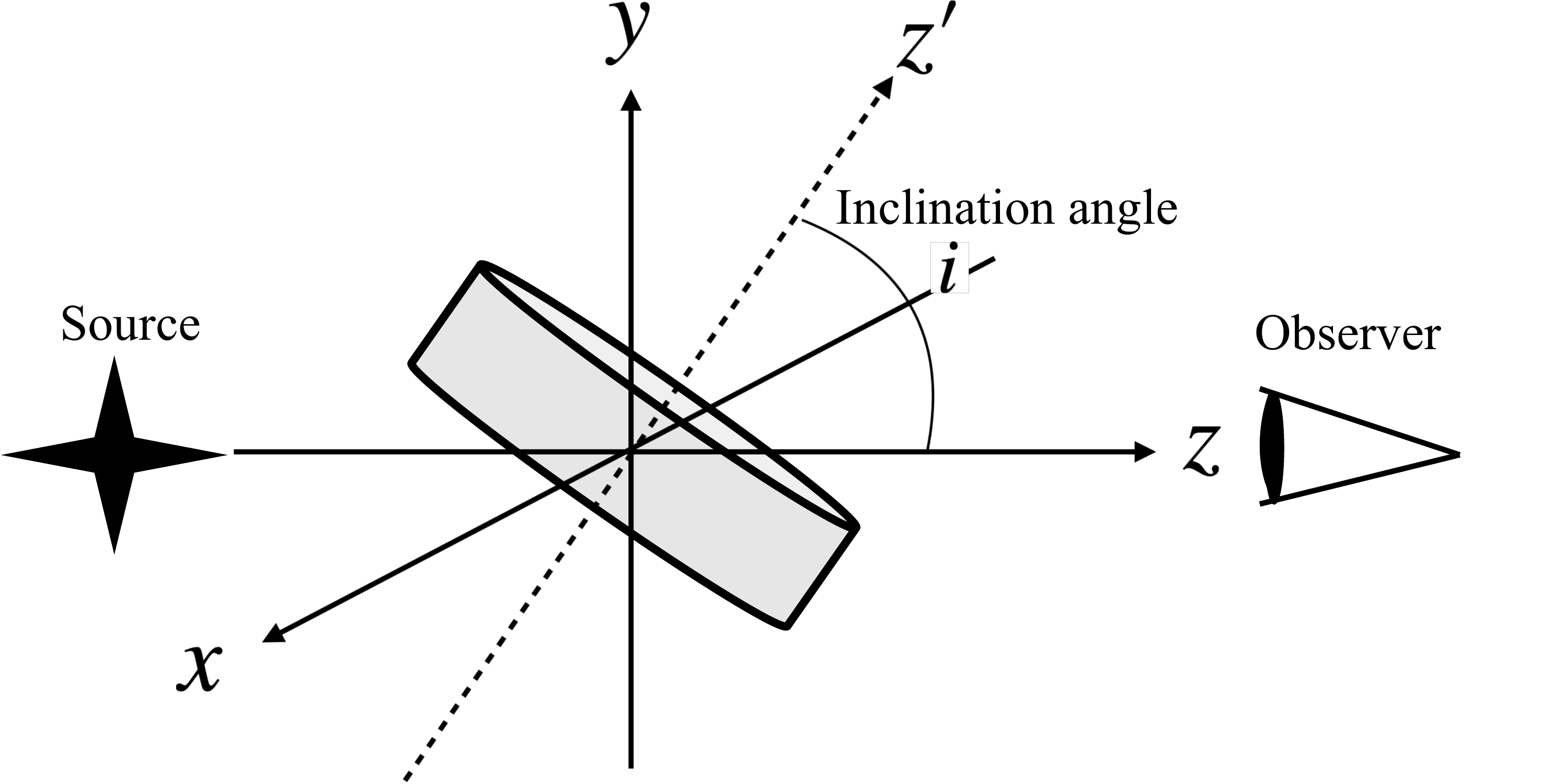}
  \end{center}
  \caption{Schematic picture of the coordinates for observation.} \label{fig:mogi}
\end{figure}

Figure \ref{fig:mogi} depicts the definition of the Cartesian coordinates in this work. We set the $z$-axis becomes the direction of the sightline and the remaining $x$-axis and $y$-axis are on the projection screen perpendicular to the $z$-axis. The vertical direction from the disk mid-plane is $z'$-axis and we define the inclination angle, $i$, as the angle from the $z$-axis to the $z^{\prime}$-axis, where $i =0^\circ$ is the face-on view and $i =90^\circ$ is the edge-on view. We place a DING at a redshift, $z$, from 0.1 to 1.0. Nearby DINGs with $z < 0.1$ are not considered because nearby galaxies are likely visible in optical and IR bands and those galaxies are no longer DING.

\begin{figure}
  \begin{center}
  	\includegraphics[width=1\linewidth,bb=0 0 596 290]{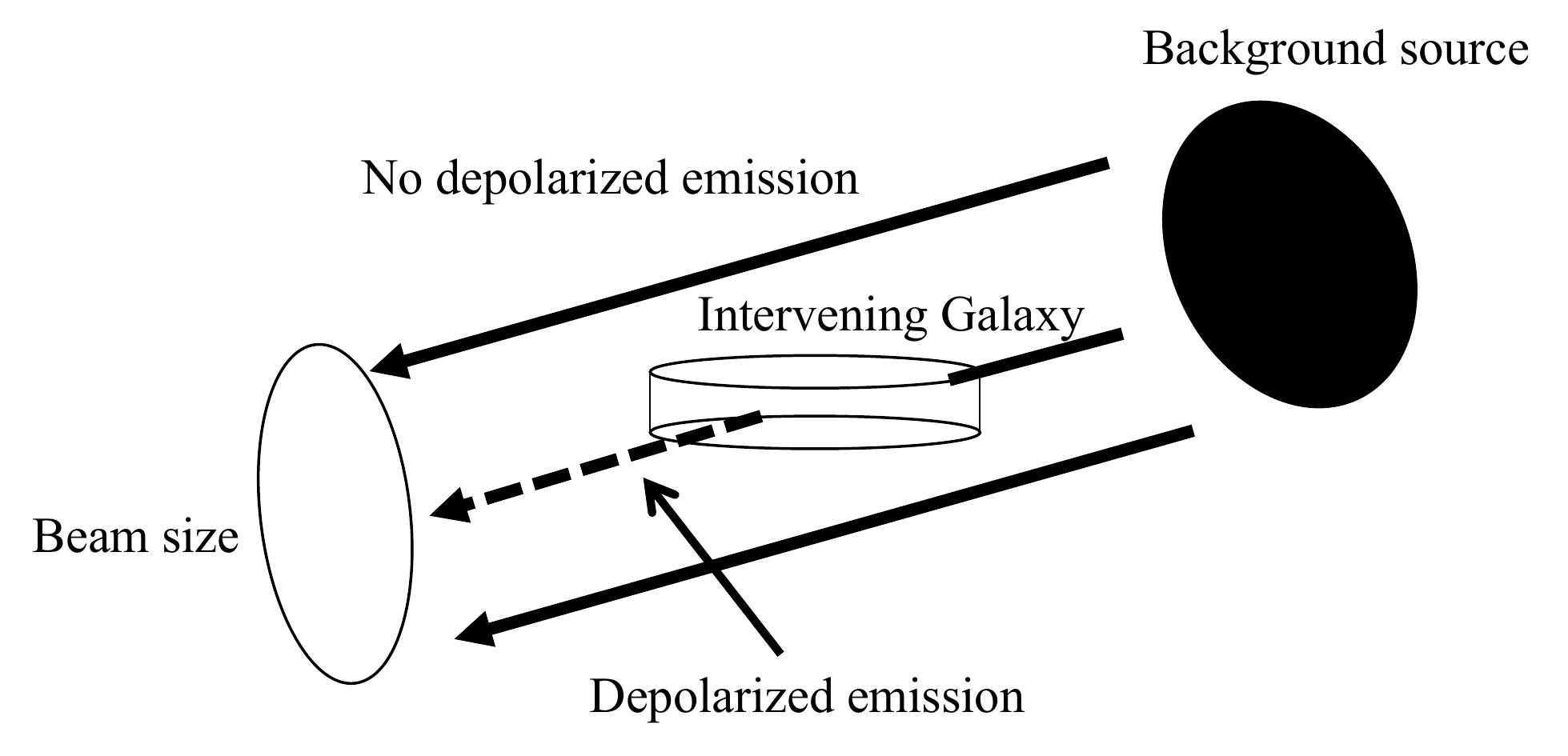}
  \end{center}
  \caption{
  A schematic picture of observed polarized emission in situation of right panel of figure \ref{fig:RMz1_x5y0_DP}}\label{fig:Depo}
\end{figure}

\subsection{Observing Specification and Calculation}

As described above, we assume that the background source has diffuse, uniform emission of a circular beam with a 1$^{\prime\prime}$ diameter. The beam passes through the DING, experiencing Faraday rotation by the RM of the DING (see Figure~\ref{fig:Depo}). This effect is calculated as follow. The DING is produced in a cubic box consisting of $200\times 200\times 200$ computational grids. The grid size, which determines the spatial resolution of the calculation, is 100 pc which is sufficiently smaller than the curvature scale ($\sim$kpc) of the disk magnetic field. The RM and Faraday rotation within the $k$-th grid along a certain line of sight are given by
\begin{eqnarray}
  \left\{
    \begin{array}{l}
    \Delta{\rm RM}_{k} = 0.81n_e B_{k,\parallel}\Delta l\\
    \Delta \chi = {\rm RM}_{k}\lambda^2\Delta l\\
    \end{array}
  \right.
\end{eqnarray}
where $B_{k,\parallel}$ is the magnetic field parallel to the line of sight, $\Delta l = 100$ pc is the line element, and $\Delta \chi$ is the Faraday rotation of the polarization angle. We integrate RM and Faraday rotation along the $z$-axis and project the cumulative RM and Faraday rotation on the $x$-$y$ plane as follows. 
\begin{eqnarray}\label{eqRM_stokes}
  \left\{
    \begin{array}{l}
      {\rm RM} = \sum_{k=1}^N \Delta{\rm RM}_k \\
      \chi = \chi_0 + {\rm RM}\lambda^2\\
      Q=Q_{0}\cos{\chi} - U_{0}\sin{\chi} \\
      U=Q_{0}\sin{\chi} + U_{0}\cos{\chi}.\\
    \end{array}
  \right.
\end{eqnarray}
We sum up $Q$ (or $U$) for the projected grids within the beam and obtain the observed $Q$ (or $U$) within the beam. The observed polarization fraction, $p$, is calculated by using the observed $Q$ and $U$ as $p = \sqrt{Q^2+U^2}/I$. The observed polarization fraction is written as $p \propto \nu^{\alpha_I-\alpha_P}=\nu^\beta$, so $\beta$ will change if the depolarization takes place. We perform the least-square-fits for the $p$-$\nu$ relation using the power-law form to estimate $\beta$.

\begin{figure}
  \begin{center}
  	\includegraphics[width=1\linewidth,bb=0 0 576 432]{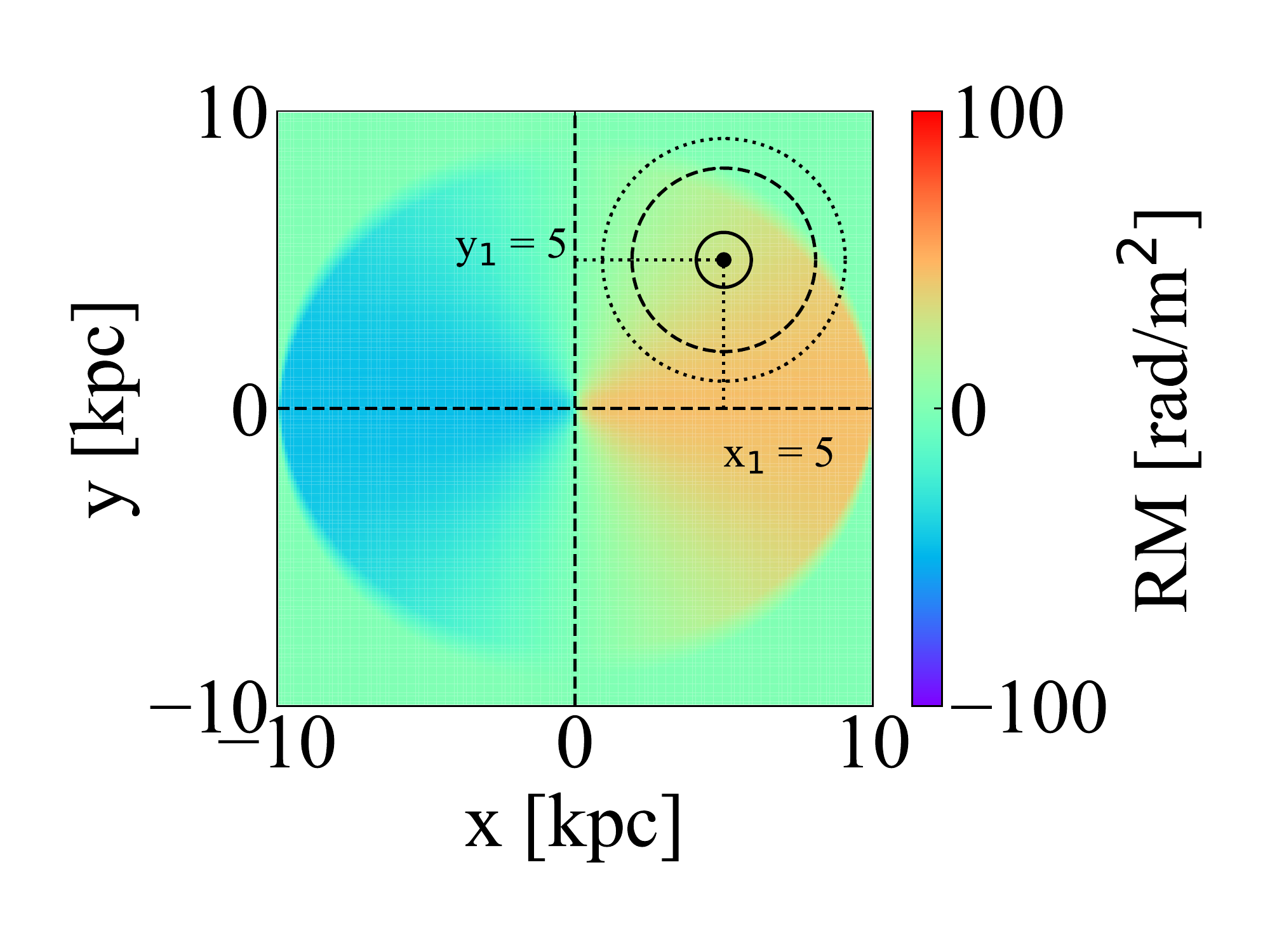}
  \end{center}
  \caption{Example of the rest-frame intrinsic RM of a DING with inclination angle $i=30^{\circ}$. The solid, dashed, and dotted circles display the references of a 1$^{\prime\prime}$ beam for  $z = 0.1$, $0.5$ and $1.0$, respectively, centered at $x_1 = 5$ kpc and $y_1 = 5$ kpc.}
\label{fig:RM}
\end{figure}

Figure \ref{fig:RM} shows the distributions of rest-frame intrinsic RM of the DING for the inclination angle $i=30^{\circ}$. We define the beam offset parameters from the center of the DING, $10\ \mathrm{kpc} \leq x_1  \leq 10\ \mathrm{kpc}$ in $x$-axis and $10\ \mathrm{kpc} \leq y_1  \leq 10\ \mathrm{kpc}$ in $y$-axis. The solid, dashed, and dotted circles display the references of a 1$^{\prime\prime}$ beam for  $z = 0.1$, $0.5$ and $1.0$, respectively, centered at $x_1 = 5$ kpc and $y_1 = 5$ kpc.

In addition to the numerical integration of RM, we derive the observed $\rm{RM}$, ${\rm RM_{peak}}$, using Faraday RM synthesis (or Faraday Tomography;  \cite{1966MNRAS.133...67B, 2005A&A...441.1217B}). We define ${\rm RM_{peak}}$ as the Faraday depth at which the amplitude of the cleaned Faraday dispersion function (FDF) is maximum, where the RM CLEAN, a technique that removes the side lobes of the dirty FDF, is used (\cite{2009IAUS..259..591H}). Such a way of RM measurement will be considered in future large surveys of extragalactic RMs such as the SKA. We prepare the sufficiently-large number of frequency channels so as to avoid numerical errors regarding a lack of channels. Hence, in the Fourier transform for Faraday tomography we use the data with the sufficiently-large number of  $\lambda^2$ channels, ensuring an ideal rotation measure spread function (RMSF). Any effects caused by coarse and/or unevenly-sampled data for Faraday tomography is beyond the scope of this paper, although the effects can be crucial in actual observational work \citep{2016PASJ...68...44M,2019MNRAS.482.2739M}. The Frequency coverage and the corresponding full width at half maximum (FWHM) of the RMSF, i.e. the resolution of Faraday depth, is listed in table \ref{band}.

We also estimate the RM using the classical $\chi - \lambda^2$ relation (see equation \ref{eqRM_stokes}), ${\rm RM_{cls}}$, and compare it with ${\rm RM_{peak}}$. Because the $\chi - \lambda^2$ relation does not follow a linear relation because of depolarization (e.g., \cite{1998MNRAS.299..189S}), it is in general difficult to derive ${\rm RM_{cls}}$ from such a polarization spectrum. We attempt to derive ${\rm RM_{cls}}$ using $\chi - \lambda^2$ relation at the relatively-high frequency, 3000 -- 4000 MHz, at which we see a nearly-linear relation for our simulations.

We perform Monte-Carlo simulations and investigate the statistical contribution of DINGs on background polarization. We set the beam offsets, $x_1$ and $y_1$, and each inclination angle from 0 degree to 90 degree with a 5 degree step as the free parameters of Monte-Carlo simulations and calculate the FDF for each redshift of 0.1, 0.3, 0.5, 0.75, and 1.0. We performed 100,000 realizations for each redshift. The convergence of the results was confirmed from the runs with different realization numbers. We adopt ${\rm RM_{peak}}$ as the observed RM for each background source and measure the DING's RM from the shift of the ${\rm RM_{peak}}$ from zero because we do not consider any other RM contribution in the simulation. We consider the 700-1800 MHz band for Faraday tomography but ${\rm RM_{peak}}$ does not significantly depend on the frequency coverage (see discussion).

\begin{table}
\tbl{Frequency coverage and FWHM in this work.}{%
\centering
  \begin{tabular}{lcc} \hline Band & Frequency\footnotemark[$^\dagger$] & FWHM\\
  & [MHz] & [$\mathrm{rad/m^2}$]\\
  \hline \hline
  Low & 150 -- 700 & 0.9098\\
  Mid & 700 -- 1800 & 22.81\\
  High & 1800 -- 4000 & 162.5 \\
  \hline
  \end{tabular}}
    \begin{tabnote}
    \footnotemark[$^\dagger$]The frequency bin is sufficiently narrow to minimize the resolution dependence.
    \end{tabnote}\label{band}
\end{table}

\section{Result}\label{Result}
\subsection{Depolarization caused by a DING}\label{result1}

\begin{figure*}
  \begin{center}
    \begin{tabular}{c}

      % 1
      \begin{minipage}{0.33\hsize}
        \begin{center}
         \includegraphics[width=1\textwidth,bb=0 0 576 432]{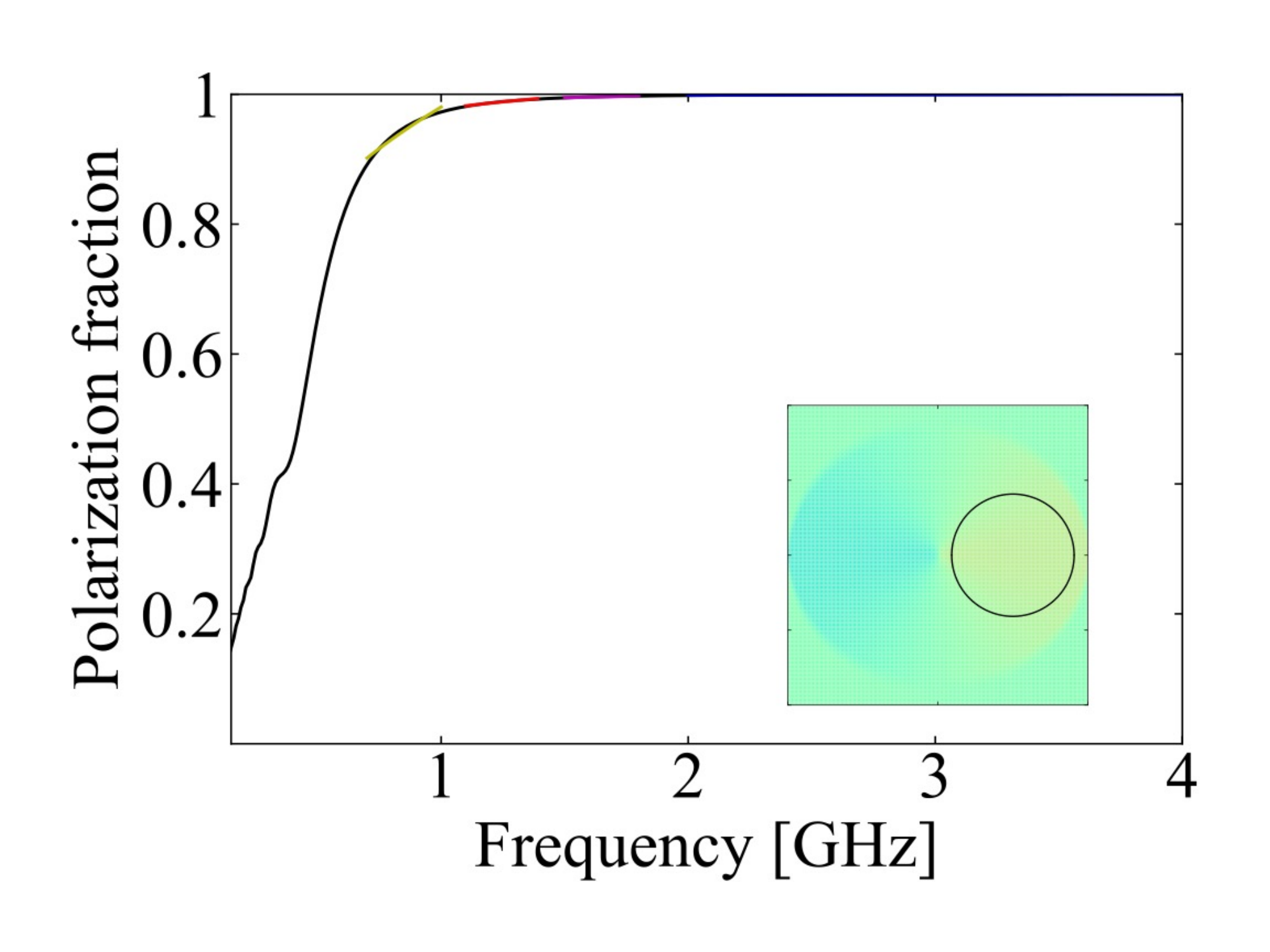}
         %\hspace{1.6cm} 
        \end{center}
      \end{minipage}
    
          % 1
      \begin{minipage}{0.33\hsize}
        \begin{center}
         \includegraphics[width=1\textwidth,bb=0 0 576 432]{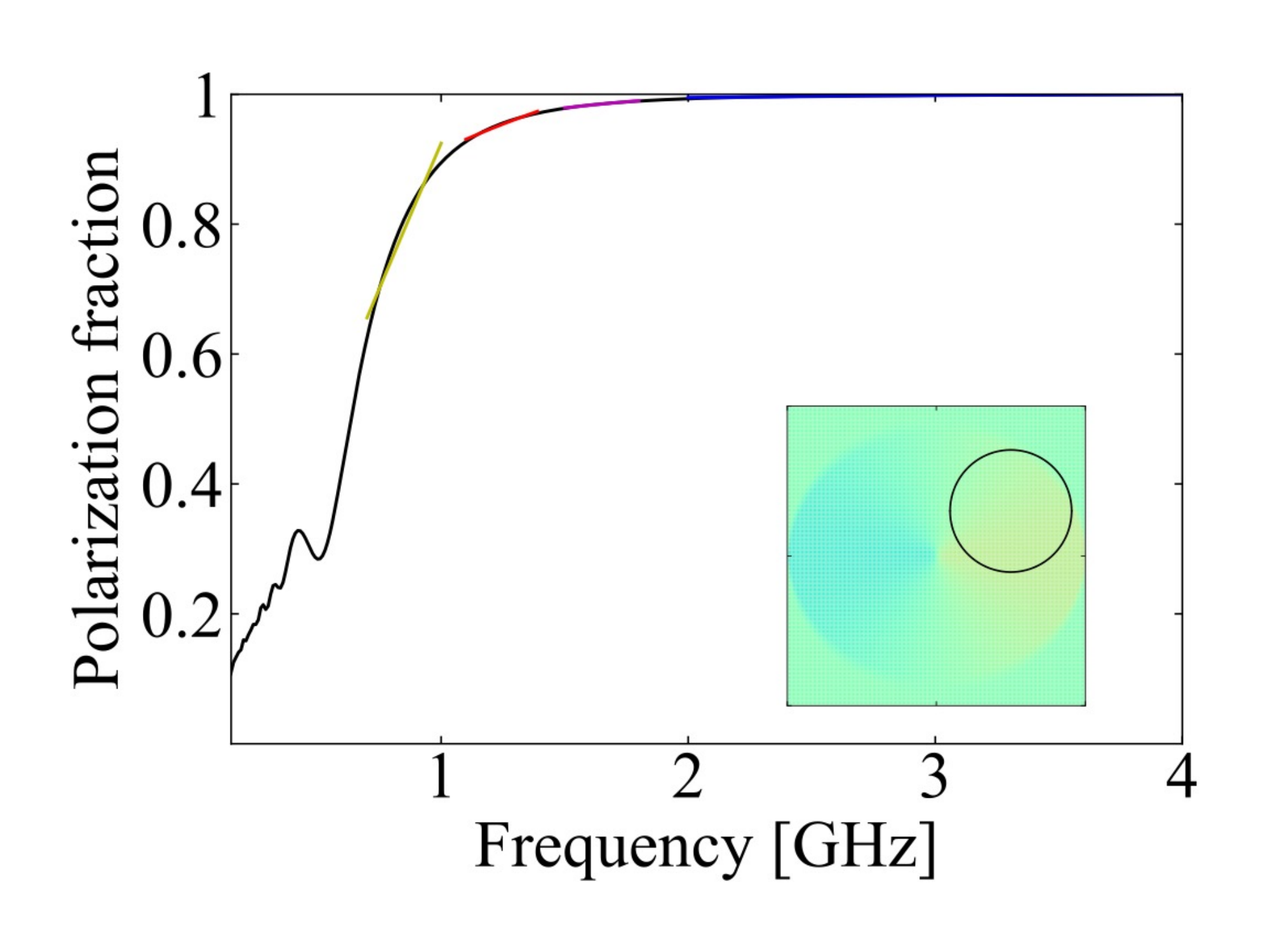}
        \end{center}
      \end{minipage}\\

      % 2
      \begin{minipage}{0.33\hsize}
        \begin{center}
          \includegraphics[width=1\textwidth,bb=0 0 576 432]{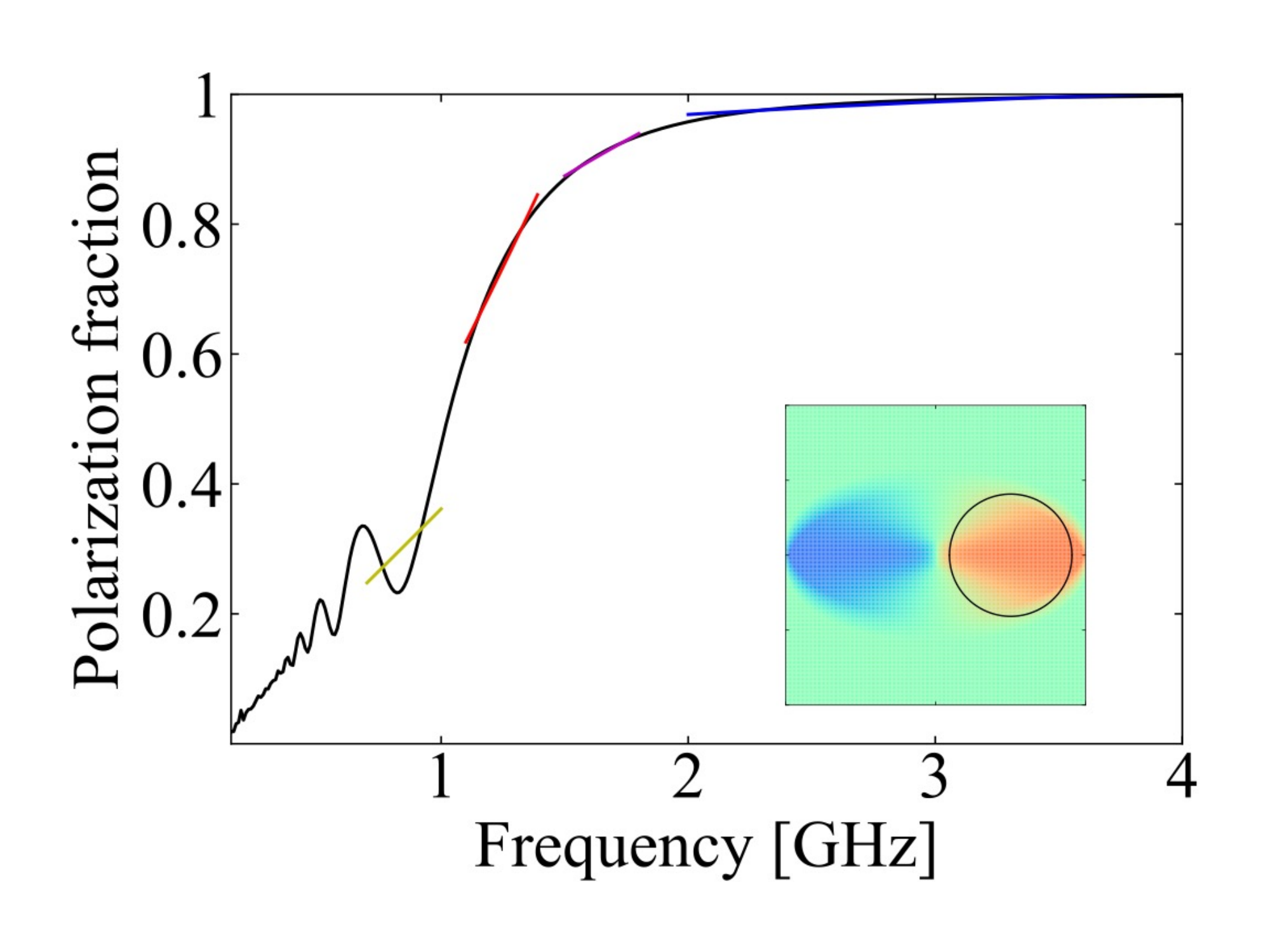}
          %\hspace{1.6cm} $i=60^{\circ}$
        \end{center}
      \end{minipage}
      
      % 2
      \begin{minipage}{0.33\hsize}
        \begin{center}
          \includegraphics[width=1\textwidth,bb=0 0 576 432]{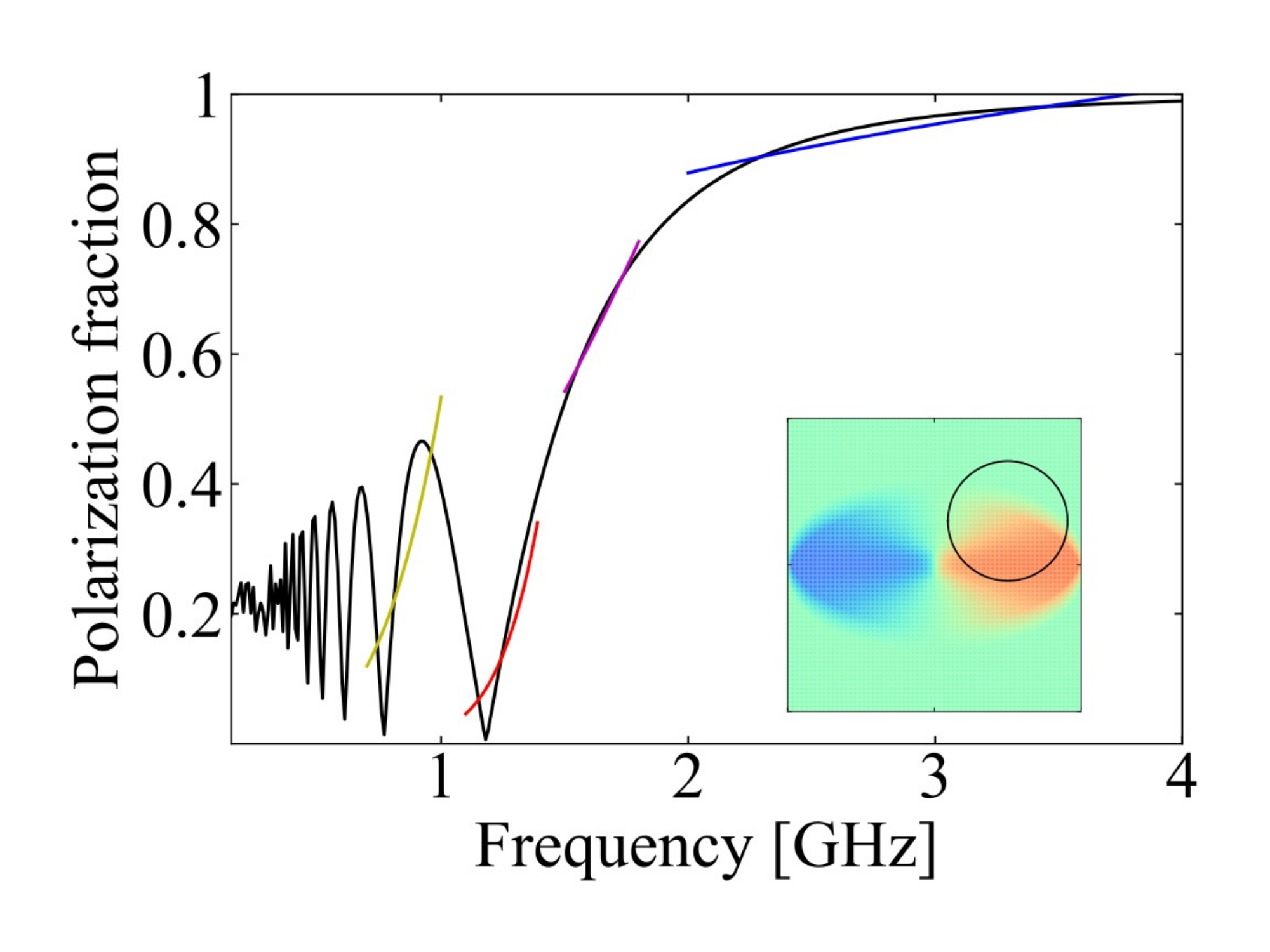}
          %\hspace{1.6cm} $i=60^{\circ}$
        \end{center}
      \end{minipage}\\

      % 3
      \begin{minipage}{0.33\hsize}
        \begin{center}
          \includegraphics[width=1\textwidth,bb=0 0 576 432]{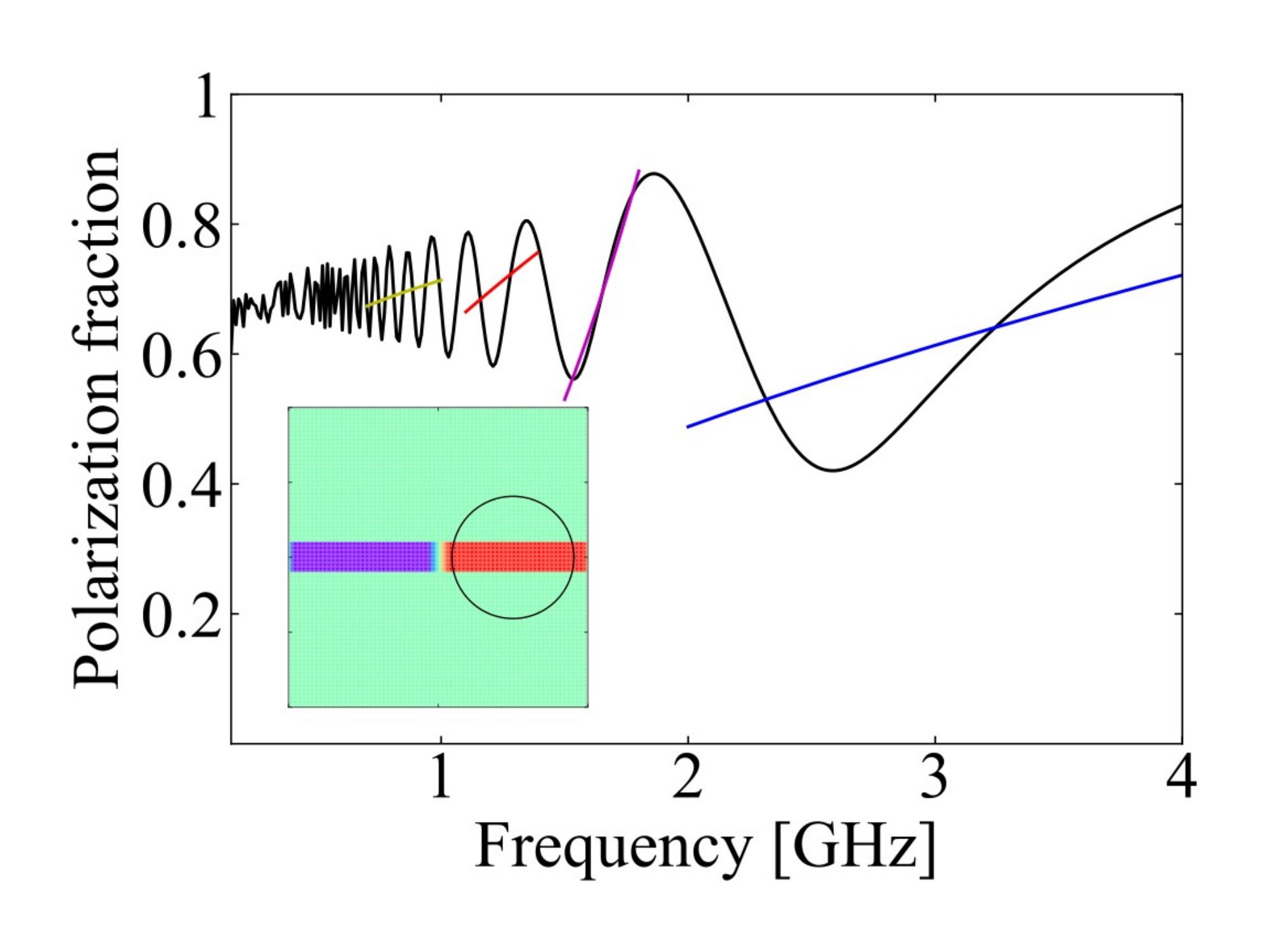}
          \hspace{1.6cm} $(z,x_1,y_1)=(1.0,5,0)$
        \end{center}
      \end{minipage}
      
      % 3
      \begin{minipage}{0.33\hsize}
        \begin{center}
          \includegraphics[width=1\textwidth,bb=0 0 576 432]{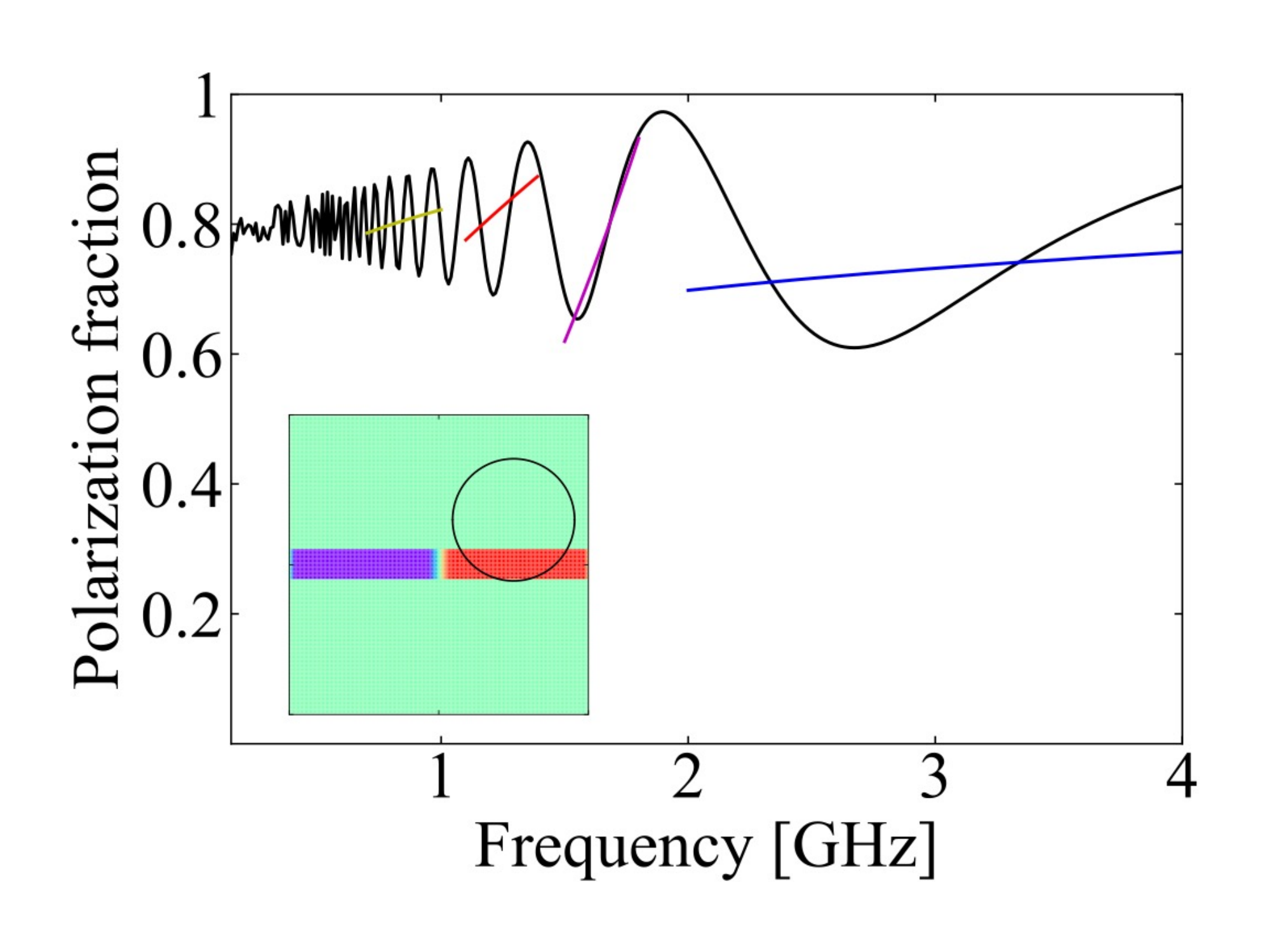}
          \hspace{1.6cm} $(z,x_1,y_1)=(1.0,5,3)$
        \end{center}
      \end{minipage}

    \end{tabular}
    \end{center}
    \caption{The polarization fraction as a function of the frequency. The left and right panels show the cases for $(z,x_1,y_1)=(1.0,5,0)$ and $(z,x_1,y_1)=(1.0,5,3)$, respectively, where the insets show the RM maps and the 1$^{\prime\prime}$ beam position is shown as the black circle. The color scale is the same as that in Figure \ref{fig:RM}. The top and bottom panels show the results for the inclination angles of 30$^\circ$, 60$^\circ$, and 90$^\circ$, respectively. The yellow, red, magenta and blue lines indicate the best fits assuming $p \propto \nu^\beta$ for the four different bands of 700$-$1000 MHz, 1100$-$1400 MHz, 1500$-$1800 MHz, and blue 2000$-$4000 MHz, respectively.}
    \label{fig:RMz1_x5y0_DP}
\end{figure*}

\begin{table}
\tbl{The average and the standard deviation of the observer-frame, in-beam RMs of the DING for $z=1.0$ and $x_1=5$~kpc.}{%
\centering
  \begin{tabular}{cccc} \hline
$y_1$ & $i$ & average & standard deviation \\
(kpc) & (degree) & ( $\mathrm{rad/m^2}$) & ( $\mathrm{rad/m^2}$) \\
 \hline \hline
0 & 30 & 10  & 1.3 \\
0 & 60 & 26 & 6.5 \\
0 & 90 & 35 & 54 \\
\hline
3 & 30 & 8.7 & 2.6 \\
3 & 60 & 16 & 13 \\
3 & 90 & 25 & 50 \\
\hline
  \end{tabular}}\label{table_av_sd}
\end{table}

Figure \ref{fig:RMz1_x5y0_DP} shows the frequency dependence of the polarization fraction for a different inclination angle and a different beam offset. The left and right panels show the results with $(z,x_1,y_1)=(1.0,5,0)$ and $(z,x_1,y_1)=(1.0,5,3)$, respectively, and the top to bottom panels show the results for the inclination angles of 30$^\circ$, 60$^\circ$, and 90$^\circ$, respectively. The insets show the RM maps, where the black circles indicate the 1$^{\prime\prime}$ beam. The average and the standard deviation of RM within the beam are summarized in table~\ref{table_av_sd}. They monotonically increase with increasing the inclination angle and decrease with increasing the beam offset.

In Figure \ref{fig:RMz1_x5y0_DP}, we see that background polarization is depolarized at low frequencies. However, although the standard deviation of RM monotonically increases with increasing the inclination angle, the degree of depolarization at low frequency does not. We see the decaying oscillation of the polarization fraction and its convergence at a certain value depending on $i$ and $y_1$. The color lines in Figure \ref{fig:RMz1_x5y0_DP} show what we obtain from the linear fit for such an oscillated profile.

Comparison between right and left panels indicates that the profile of the polarization fraction significantly depends on the beam offset. A decaying oscillation caused by depolarization appears at a smaller inclination angle (see the results for $i=60^\circ$) as we increase the beam offset. Since the beam offset determines the filling factor or the covering fraction of the DING for a beam, the filling factor is an essential parameter. Actually, for $(y_1, i)=(3, 60^{\circ})$, the polarization fraction converges at about 0.2 because the filling factor is about 0.8. The profile of the polarization fraction for $(y_1, i)=(3, 60^{\circ})$ is similar to that for $(y_1, i)=(0, 60^{\circ})$, but the polarization fraction of the former converges at a higher value than that of latter, because the polarized emission passes through more $0\ \mathrm{rad/m^2}$ regions. As for $i=30^{\circ}$, we do not see the convergence of the polarization fraction. This is reasonable because the DING fully covers the beam area and the filling factor is unity. Similar results are found for $x_1$ instead of $y_1$.

The ${\rm RM_{cls}}$ derived from the $\chi - \lambda^2$ relation within the beam are summarized in table~\ref{table_classRM}. ${\rm RM_{cls}}$ are mostly identical to the beam-averaged RM for $i=30^{\circ}$ and $i=60^{\circ}$, while they are very different from each other for $i=90^{\circ}$. The difference is made by significant depolarization which is seen even at the high frequency of 4 GHz (see Figure \ref{fig:RMz1_x5y0_DP}). The $\chi - \lambda^2$ relation suffers from estimating the beam-averaged RM in such a strongly-depolarized spectrum.

\begin{table}
\tbl{The observed RM, ${\rm RM_{cls}}$, with the $\chi - \lambda^2$ relation in-beam RMs of the DING for $z=1.0$ and $x_1=5$~kpc.}{%
\centering
  \begin{tabular}{ccc} \hline
$y_1$ & $i$ & ${\rm RM_{cls}}$ \\
(kpc) & (degree) & ( $\mathrm{rad/m^2}$) \\
 \hline \hline
0 & 30 & 10\\
0 & 60 & 26\\
0 & 90 & 12\\
\hline
3 & 30 & 8.7\\
3 & 60 & 16\\
3 & 90 & -0.85\\
\hline
  \end{tabular}}\label{table_classRM}
\end{table}

\subsection{Estimation of DING's RM}\label{result2}

\begin{figure*}
  \begin{center}
    \begin{tabular}{c}

      % 1_M
      \begin{minipage}{0.33\hsize}
        \begin{center}
         \includegraphics[width=1\textwidth,bb=0 0 576 433]{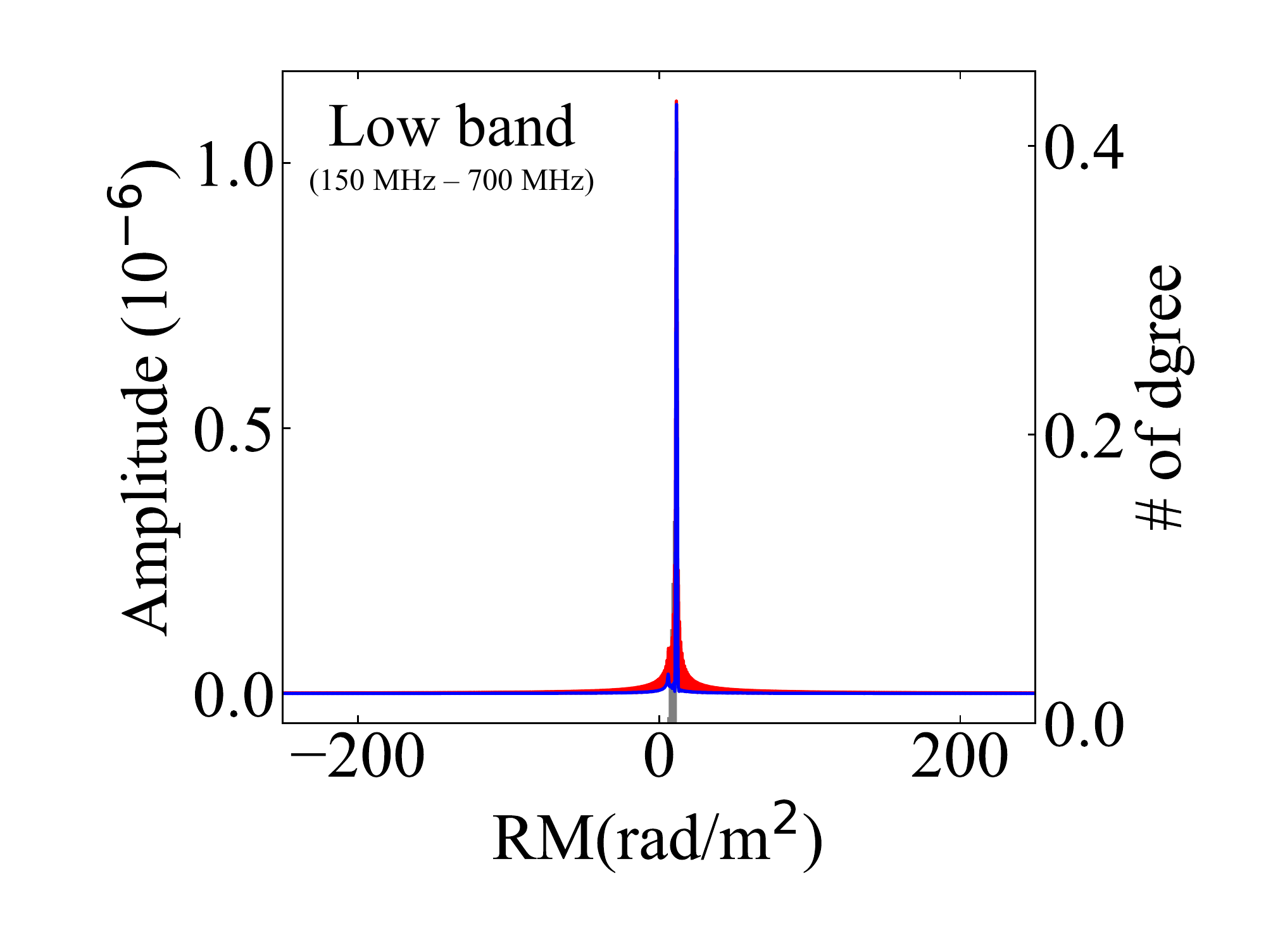}
        \end{center}
      \end{minipage}

      % 2_M
      \begin{minipage}{0.33\hsize}
        \begin{center}
          \includegraphics[width=1\textwidth,bb=0 0 576 432]{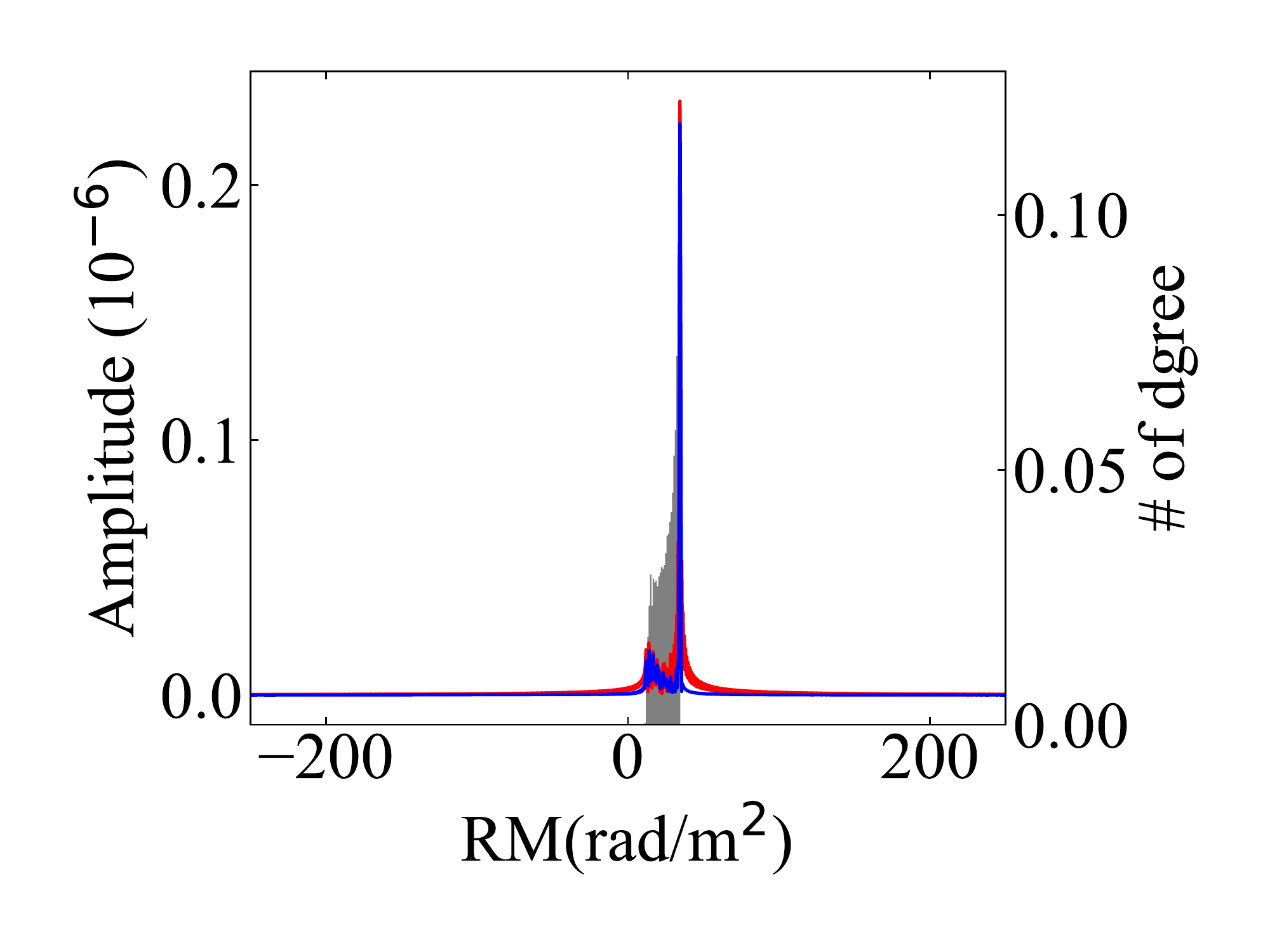}
        \end{center}
      \end{minipage}

      % 3_M
      \begin{minipage}{0.33\hsize}
        \begin{center}
          \includegraphics[width=1\textwidth,bb=0 0 576 432]{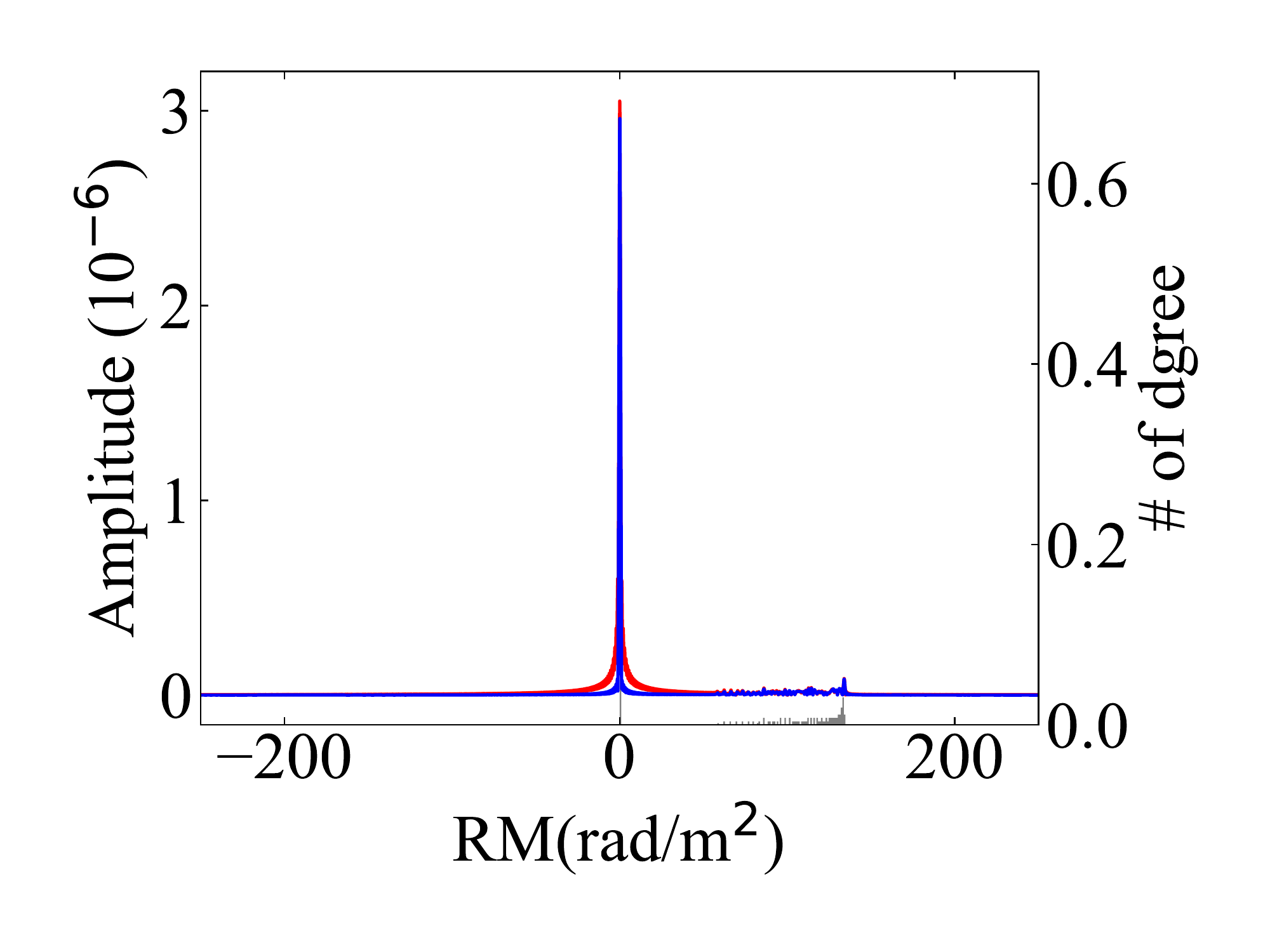}
        \end{center}
      \end{minipage}\\

      % 1_G
      \begin{minipage}{0.33\hsize}
        \begin{center}
         \includegraphics[width=1\textwidth,bb=0 0 576 433]{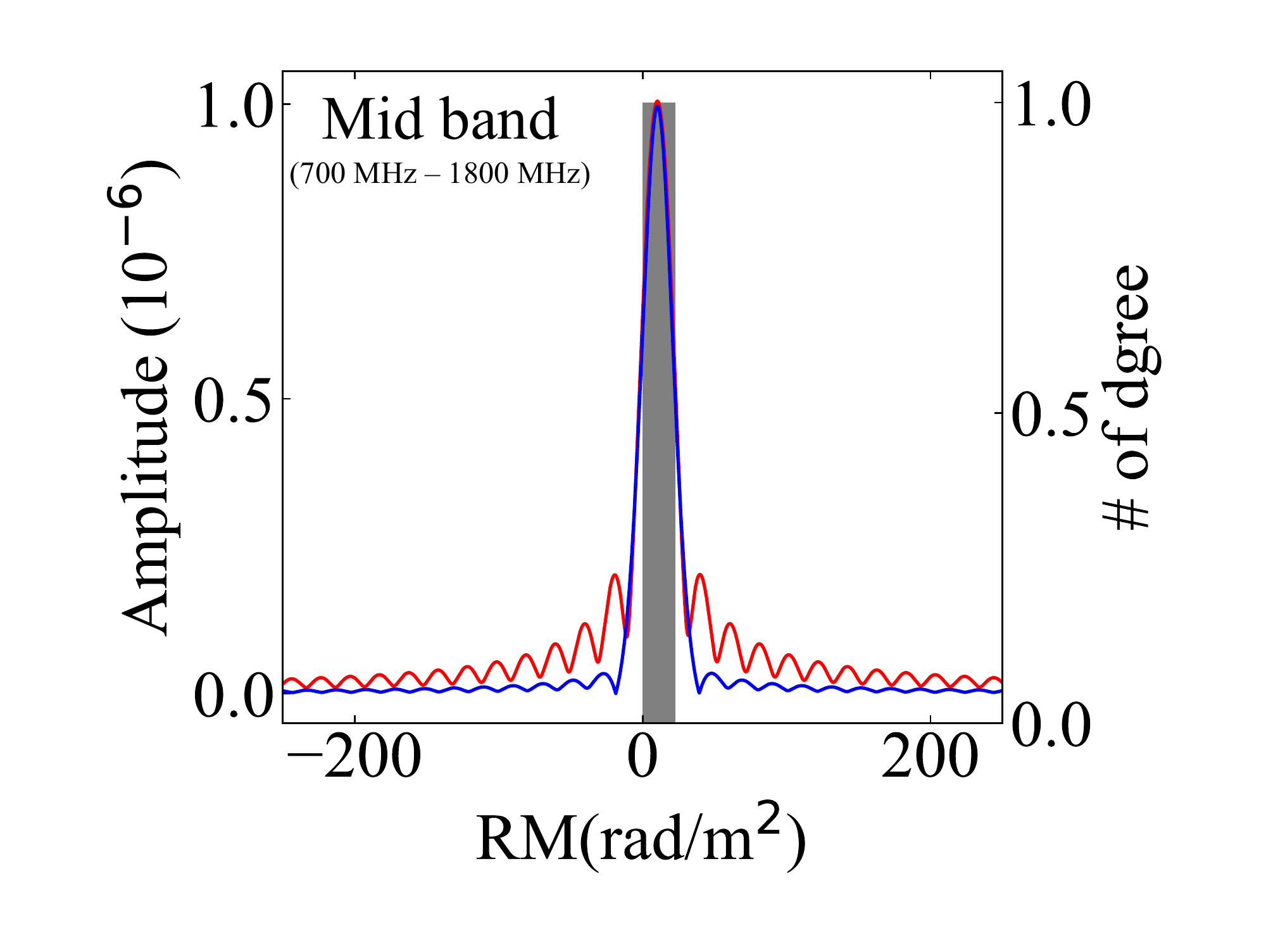}
        \end{center}
      \end{minipage}

      % 2_G
      \begin{minipage}{0.33\hsize}
        \begin{center}
          \includegraphics[width=1\textwidth,bb=0 0 576 432]{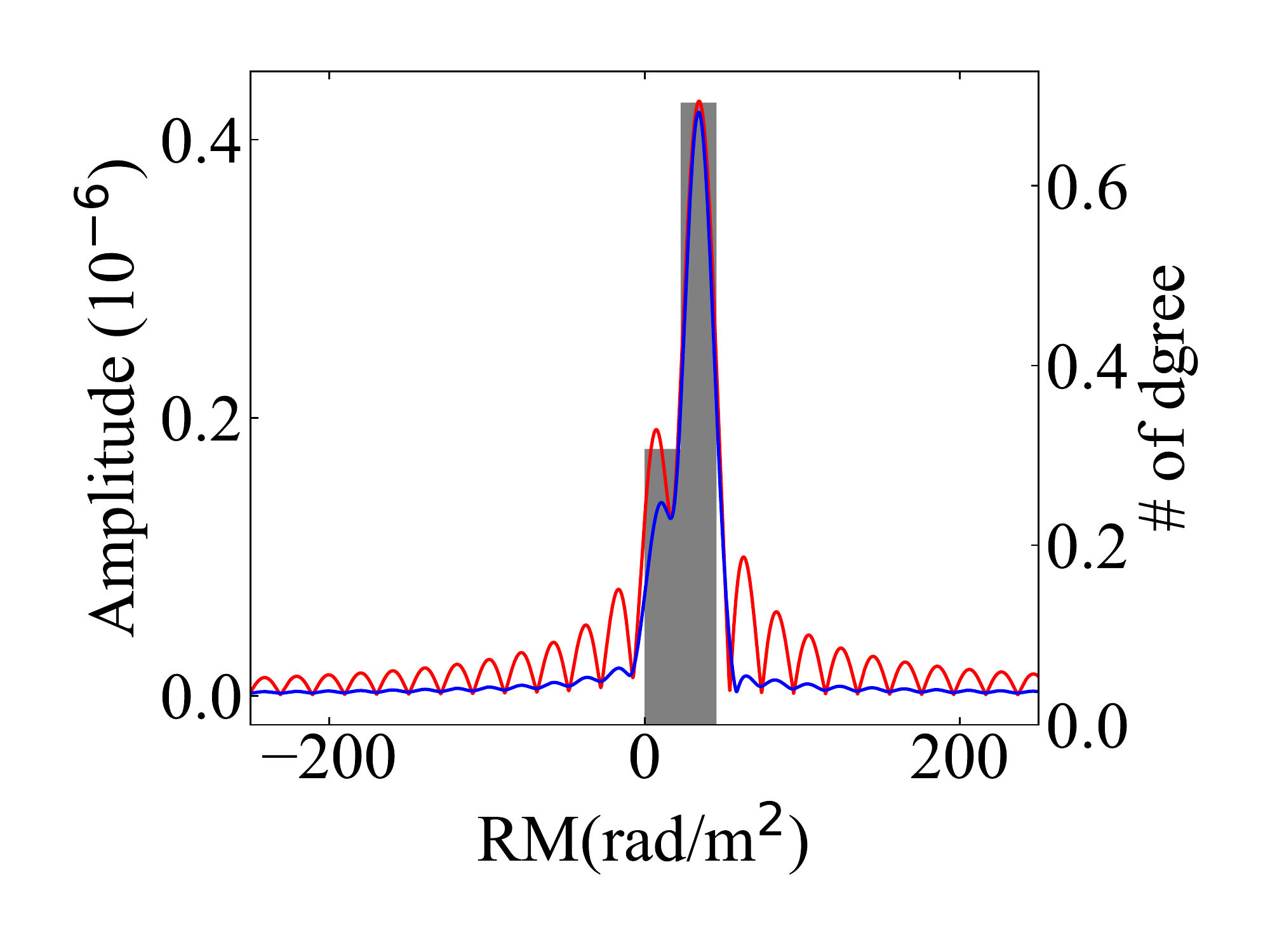}
        \end{center}
      \end{minipage}

      % 3_G
      \begin{minipage}{0.33\hsize}
        \begin{center}
          \includegraphics[width=1\textwidth,bb=0 0 576 432]{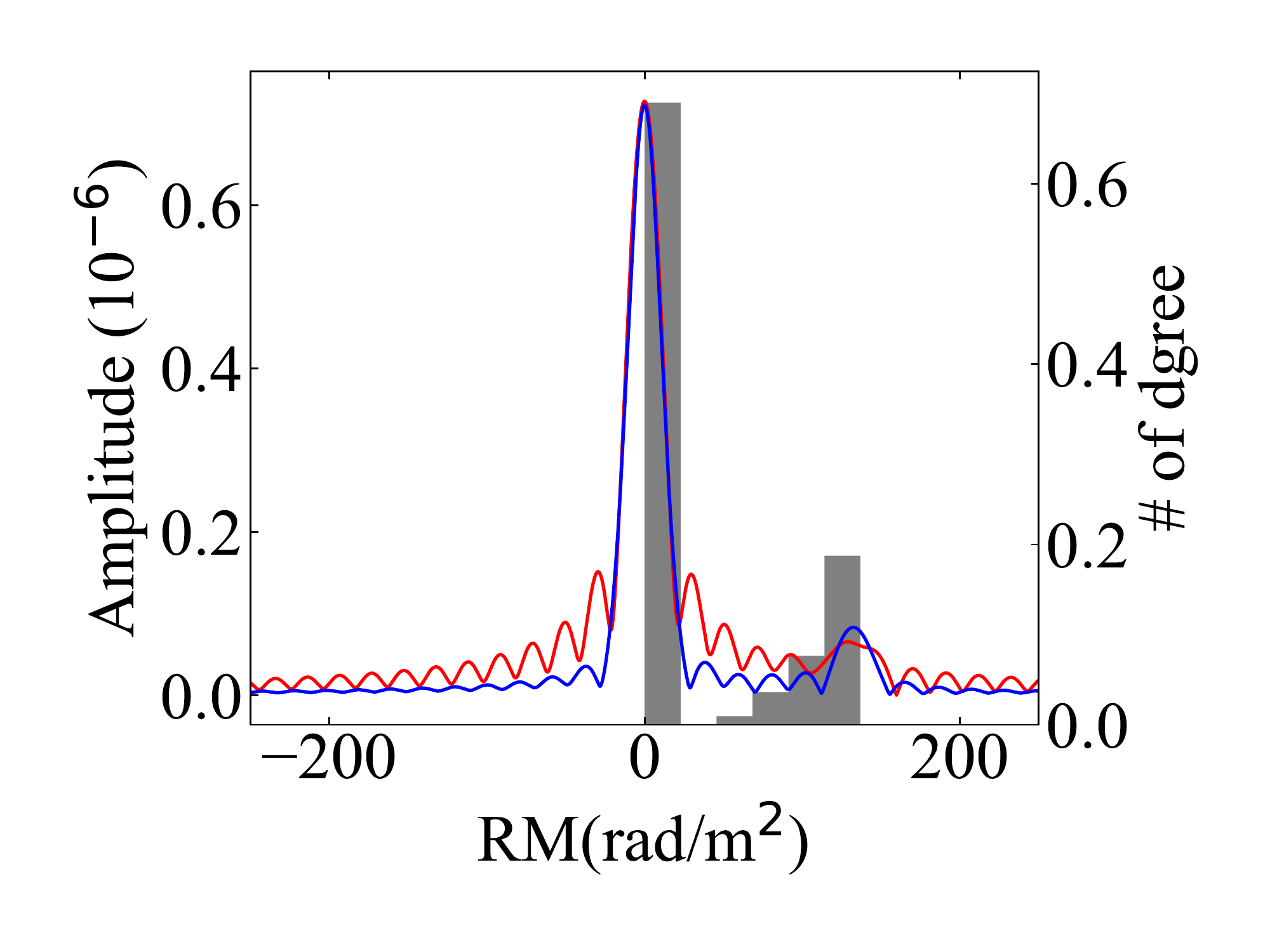}
        \end{center}
      \end{minipage}\\

      % 1_A
      \begin{minipage}{0.33\hsize}
        \begin{center}
         \includegraphics[width=1\textwidth,bb=0 0 576 433]{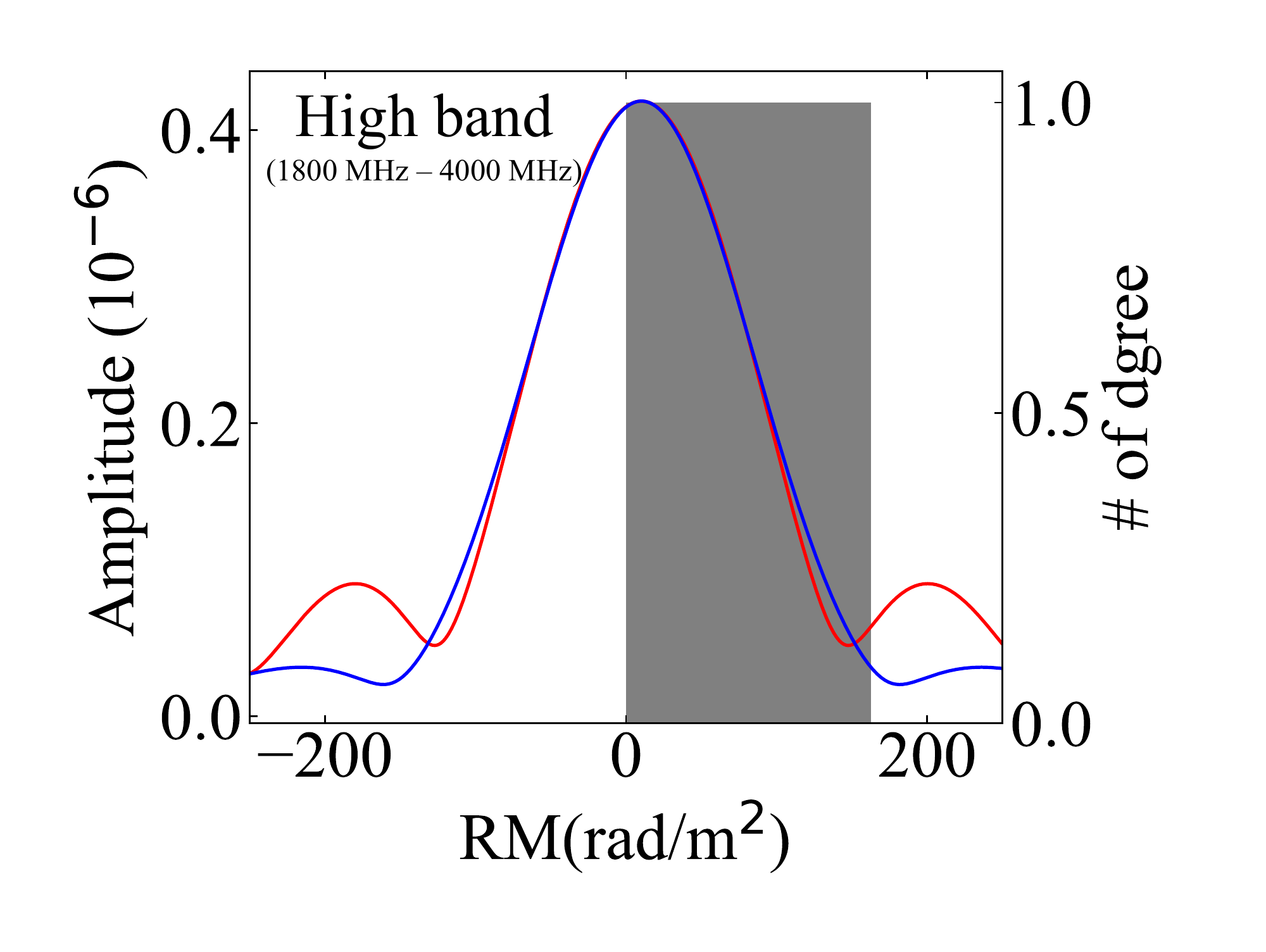}
         \hspace{1.6cm} $i=30^{\circ}$
        \end{center}
      \end{minipage}

      % 2_A
      \begin{minipage}{0.33\hsize}
        \begin{center}
          \includegraphics[width=1\textwidth,bb=0 0 576 432]{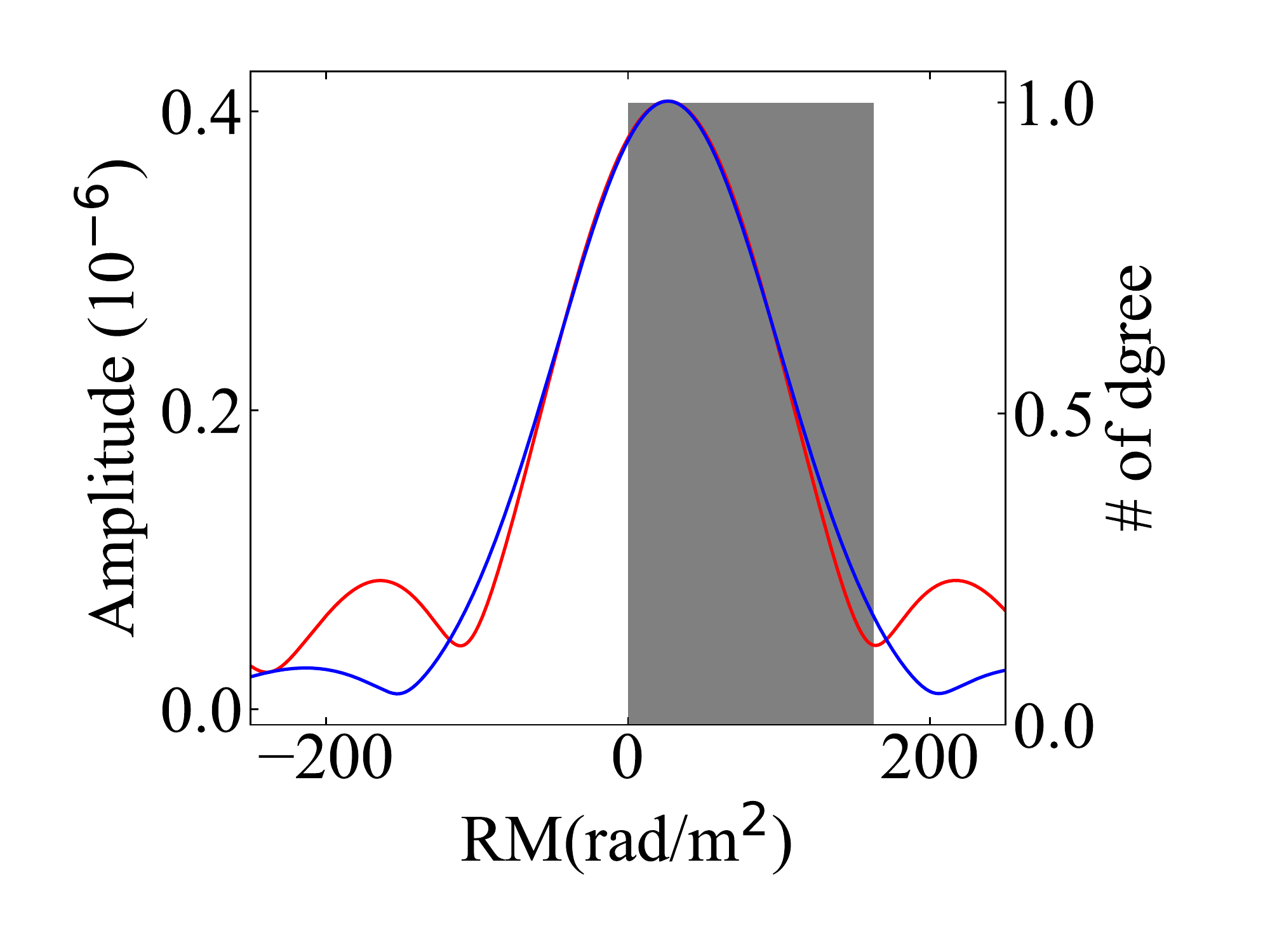}
          \hspace{1.6cm} $i=60^{\circ}$
        \end{center}
      \end{minipage}

      % 3_A
      \begin{minipage}{0.33\hsize}
        \begin{center}
          \includegraphics[width=1\textwidth,bb=0 0 576 432]{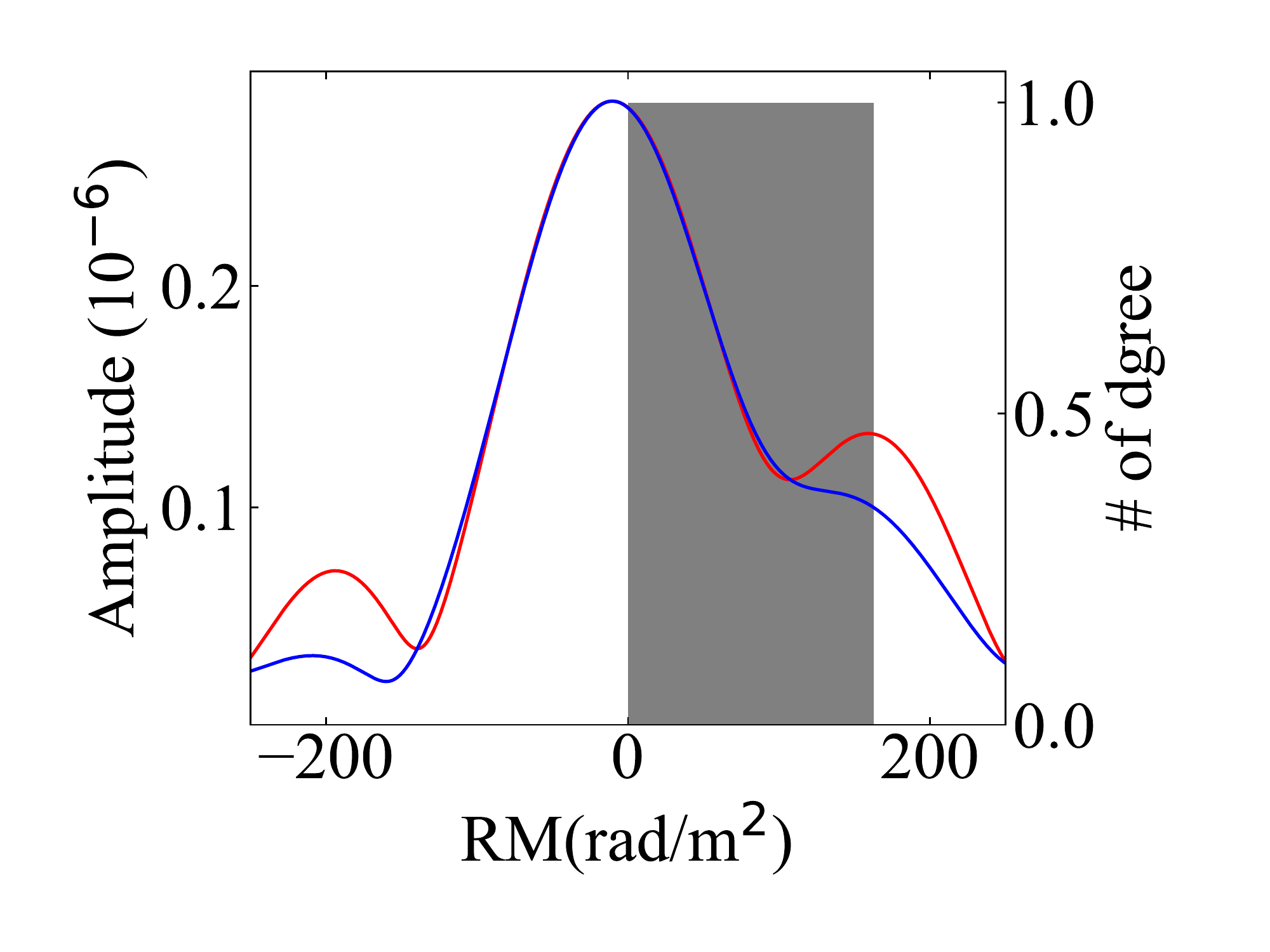}
          \hspace{1.6cm} $i=90^{\circ}$
        \end{center}
      \end{minipage}\\
    \end{tabular}
    \end{center}
    \caption{Histograms of RM (gray) and simulated FDFs (red and blue) for different inclination angles of the DING (left to right) and different frequency coverage (top to bottom). The gray color shows the histogram of RM, where the histograms are normalized by $1/(1+z)^2$. Red and blue lines show the dirty and cleaned FDFs, respectively. Panels from left to right correspond to the cases for the left to right panels shown in Figure~\ref{fig:RMz1_x5y0_DP}.}
    \label{fig:RMz1_x5y0_RMsys}
\end{figure*}

Figure \ref{fig:RMz1_x5y0_RMsys} shows the Faraday spectra obtained from Stokes parameters within the beam with $(z,x_1,y_1)=(1.0,5,0)$. Panels from the top to bottom show the results for low, mid, and high bands, respectively. The red and blue lines show the amplitudes of the dirty FDF obtained from Fourier transform of the complex polarized intensity and the cleaned FDF obtained from RM CLEAN, respectively. We also show the histograms of RM as the filled gray boxes, where the histograms are for observer-frame RMs of the projected numerical grids within the beam. If one refers to a rest-frame value of RM, it is reproduced by multiplying $(1 + z)^2$. The bin size of $x$-axis (RM or Faraday depth) is equal to the FWHM of the RMSF for each band. The histogram does not depend on the observing band, but its appearance changes according to the adopted bin size.

If there is no DING, both the FDF and the histogram should have one peak at ${\rm RM}=0$~$\mathrm{rad/m^2}$ with the zero width (i.e. the delta function). Figure \ref{fig:RMz1_x5y0_RMsys} clearly indicates non-zero values of them, confirming that the DING significantly contributes to the FDF. Interestingly, the FDF and the histogram tend to show two peaks as the inclination angle increases (see e.g., the case with Mid band for $i=90^{\circ}$). The emission passing through a larger RM of the DING appears as a peak at non-zero Faraday depth, while the emission passing through a smaller RM of the DING or the emission not passing through the DING form another peak around ${\rm RM}=0$~$\mathrm{rad/m^2}$. For $i=90^{\circ}$, most of the emission does not pass through the DING. Thus, the component around ${\rm RM}=0$~$\mathrm{rad/m^2}$ becomes the dominant component of the FDF. The inclination angle close to the edge-on results in a larger RM gap between the two components than the case of a face-on, which means that the gap can be found at higher frequency even with a larger FWHM. Indeed, for $i=60^{\circ}$, it is difficult to identify the two component with high band, while for $i=90^{\circ}$ the two components are seen in the FDF even with high band. We note that this component may be numerical artifacts in Fourier transform (see Section \ref{FDFdiscuss}).

We find that ${\rm RM_{peak}}$ tends to represent the RM value of the peak position of the histogram, and is clearly different from the actual, beam-averaged, observer-frame RM of the DING. The trend is clearly seen at the low band. For example, for $i=30^{\circ}$, ${\rm RM_{peak}} \sim 12\ \mathrm{rad/m^2}$ overestimates the beam-averaged RM, $\sim 10\ \mathrm{rad/m^2}$, by about 15~\%. Similarly, for $i=60^{\circ}$, ${\rm RM_{peak}} \sim 34\ \mathrm{rad/m^2}$ overestimates the beam-averaged one, $26\ \mathrm{rad/m^2}$. Then, when the undepolarized, intrinsic component dominates the FDF such as the case for $i=90^{\circ}$, ${\rm RM_{peak}}$ suggests $0.0\ \mathrm{rad/m^2}$, although the beam-averaged RM is $35\ \mathrm{rad/m^2}$. The classical estimate of RM, ${\rm RM_{cls}}$, tends to represent the mean RM of the LOS and is hence close to the beam-averaged RM value. As a result, ${\rm RM_{peak}}$ is also different from ${\rm RM_{cls}}$.

Overall, the shape of the cleaned FDF is similar to the histogram, except the result of Low band for $i=60^{\circ}$ which shows that the position of the secondary peak component is not at the peak of the histogram but around the edge of the histogram. We will discuss the reason in Section \ref{FDFdiscuss}.

\subsection{Statistical Contribution of DING}\label{statis}

Although our model of the DING is too simple to make quantitative predictions, we demonstrate how the effect of the DING qualitatively emerges (or is hidden) in statistical studies. Particularly, we focus on the redshift evolution of observed RMs, because such a study is one of the key Magnetism science of forthcoming surveys (Section \ref{intro}).

To study the statistical contribution of the DING, we carry out Monte-Carlo simulations as described in Section \ref{model_cal}. A foreground RM screen such as the DING shifts the Faraday depth of a background source, and the amount of the shift depends on the strength of DING's magnetic field as well as the DING's parameter such as the inclination angle and the beam offset. Thus, in a catalog of background polarized sources, DING's RM appears as the standard deviation of the observed RMs, $\sigma_{\rm RM}$, where the observed RM is defined as the Faraday depth at the FDF peak, RM$_{\rm peak}$, using Faraday tomography.

\begin{figure}
  \begin{center}
  	\includegraphics[width=1\linewidth,bb=0 0 576 432]{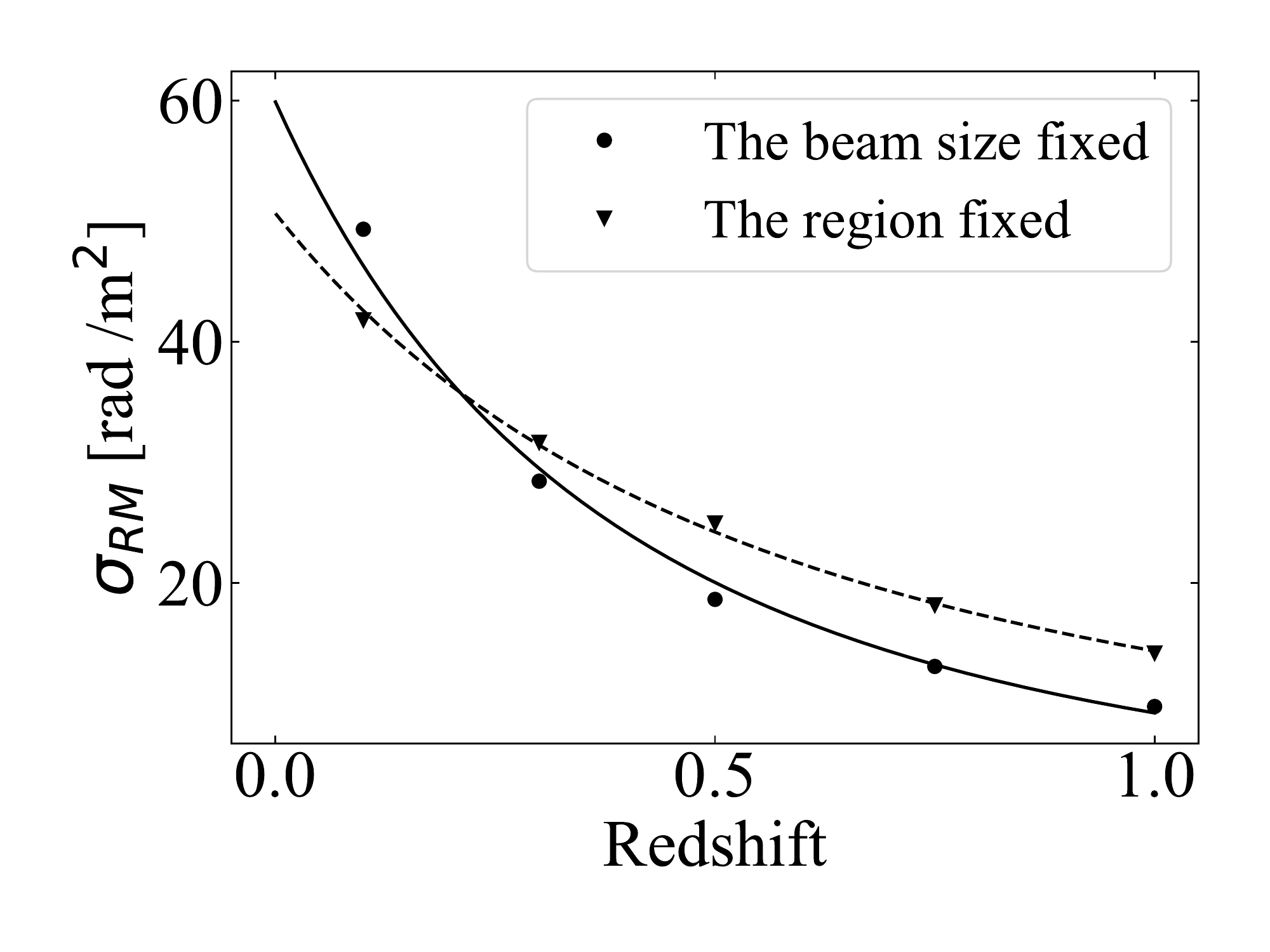}
  \end{center}
  \caption{Standard deviation of ${\rm RM_{peak}}$, $\sigma_{\rm RM}$, as a function of the redshift, $z$. The circles and triangles, respectively, show the results of our fiducial runs and our control runs for which we fix the physical beam size of 1.8 kpc in diameter. The solid and dashed lines are the best least-square fit assuming the power-law, for our fiducial and control runs, respectively.}\label{fig:MCRMsig_fit}
\end{figure}

The standard deviation of the DING's RMs as a function of the DING's redshift is shown as the circles in Figure \ref{fig:MCRMsig_fit}. We fit $\sigma_{\rm RM}$ with the following exponential function;
\begin{equation}
\label{RMsig}
\sigma_{\rm RM}(z) = \frac{A}{(1+z)^k}\ [\mathrm{rad/m^2}].
\end{equation}
The best-fit (the solid line) values are $A \sim 60$ and $k \sim 2.7$. 

The observed RM is always $1/(1+z)^2$ times the intrinsic value because RM is proportional to the square of the wavelength. To confirm this dependence, we perform the control run in which the region where the emission passes through the DING is constant regardless of the redshift (the beam size is changed at each redshift). The size of the region to be passed through is thus fixed at 1.8 kpc in diameter. The result is also shown as the triangles in Figure \ref{fig:MCRMsig_fit}. The best-fit (the dashed line) values are $A \sim 51$ and $k \sim 1.8$ which is marginally consistent with the dependence of $1/(1+z)^2$. Therefore, the steeper redshift dependence of $k \sim 2.7$ relates to the change of the physical size of the beam. We discuss this mechanism in section \ref{staticdis}.

\section{Discussion}\label{Discuss}
\subsection{Model Dependence}

We conducted numerical simulations of depolarization caused by a DING using the simplest model consisting of a uniform density and a ring magnetic field. In this subsection, we expand our simulations to more complicated but more realistic models of a DING, although the essence of depolarization effects caused by a DING can be understood based on the results of our simplest model.

\subsubsection{Density Profile}

First, the global electron distribution is assumed to be uniform in this paper. The global electron distribution has been studied in the literature and some realistic models have been proposed such as NE2001 \citep{2002astro.ph..7156C} and YMW16 \citep{2017ApJ...835...29Y}. These works suggest that the global electron distribution basically follows an exponential profile as functions of the distances from the galactic center and the disk mid-plane. Such a profile induces the two additional effects, bias and gradient. The bias effect is that RM is the largest around the central part of the DING and weaker at its outskirt. Nevertheless, the electron density is positive definite, so that the change of the sign of RM caused by the disk magnetic field remains regardless of the electron distribution model. The gradient effect is that the density gradient induces the RM gradient as well, resulting in more depolarization within the beam in general. 

To see the effect of the density profile, we adopt the exponential distribution for the electron density (Equation \ref{eq:electron_d}) on our simplest ring magnetic field model and study the same case of $(z,x_1,y_1)=(1.0,5,0)$. We then see no apparent change of the results for $i=30^\circ$ and $i=60^\circ$, because the main disk component has approximately the same density value for the uniform and exponential models. Meanwhile, we see a difference between the two models for $i=90^\circ$. The second peak of FDFs, as seen in the right column of Figure \ref{fig:RMz1_x5y0_RMsys}, looks flat, because the electron distribution is exponential.

The above result clarifies that the exponential distribution does not significantly change the depolarization through the galactic disk. Hereafter, we consider the exponential distribution of the electron density to incorporate this effect on our additional simulations shown in the following sections.

\subsubsection{Disk Fields}

We considered the ring magnetic fields which indicate one pair of positive and negative RM structures through the disk (see e.g., Figure \ref{fig:RM}). Such pairs can be obtained if we consider a spiral magnetic field such as the axi-symmetric spiral (ASS). Significant depolarization and the bi-modal FDF profile are expected, if the beam covers such the pairs producing the large RM dispersion. If we consider even higher-mode spirals such as the bi-symmetric spiral (BSS), there are more reversals of magnetic fields and multiple pairs of positive and negative RM structures through the disk. Thus, the number of the peak components in the FDF depends on the mode. The mode also changes the amount of depolarization, because the dispersion of RM within the beam depends on the mode.

\begin{figure*}
  \begin{center}
    \begin{tabular}{c}
      \begin{minipage}{0.33\hsize}
        \begin{center}
         \includegraphics[width=1\textwidth,bb=0 0 576 433]{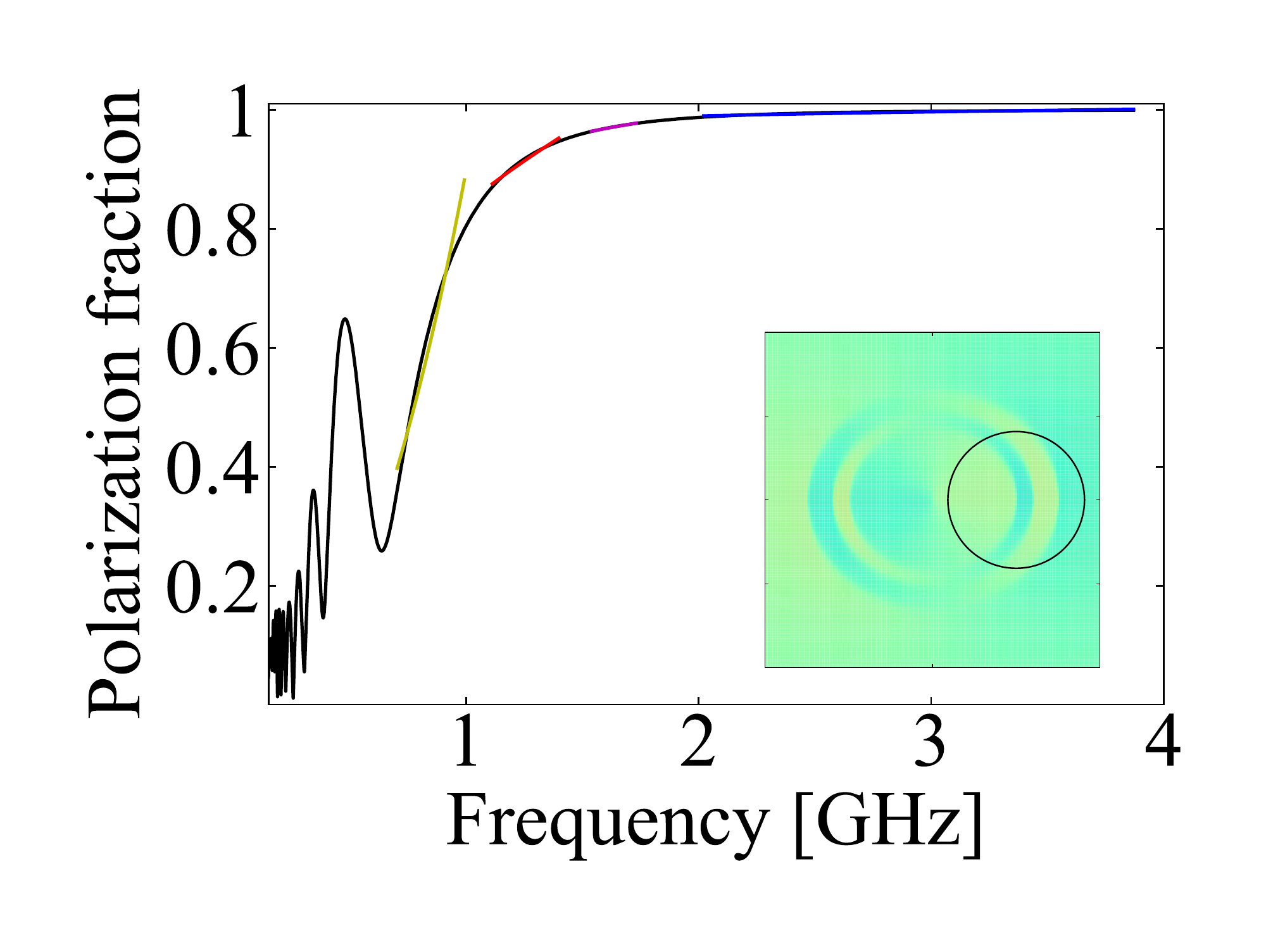}
         \hspace{0.5cm}
        \end{center}
      \end{minipage}

      % 2_A
      \begin{minipage}{0.33\hsize}
        \begin{center}
          \includegraphics[width=1\textwidth,bb=0 0 576 432]{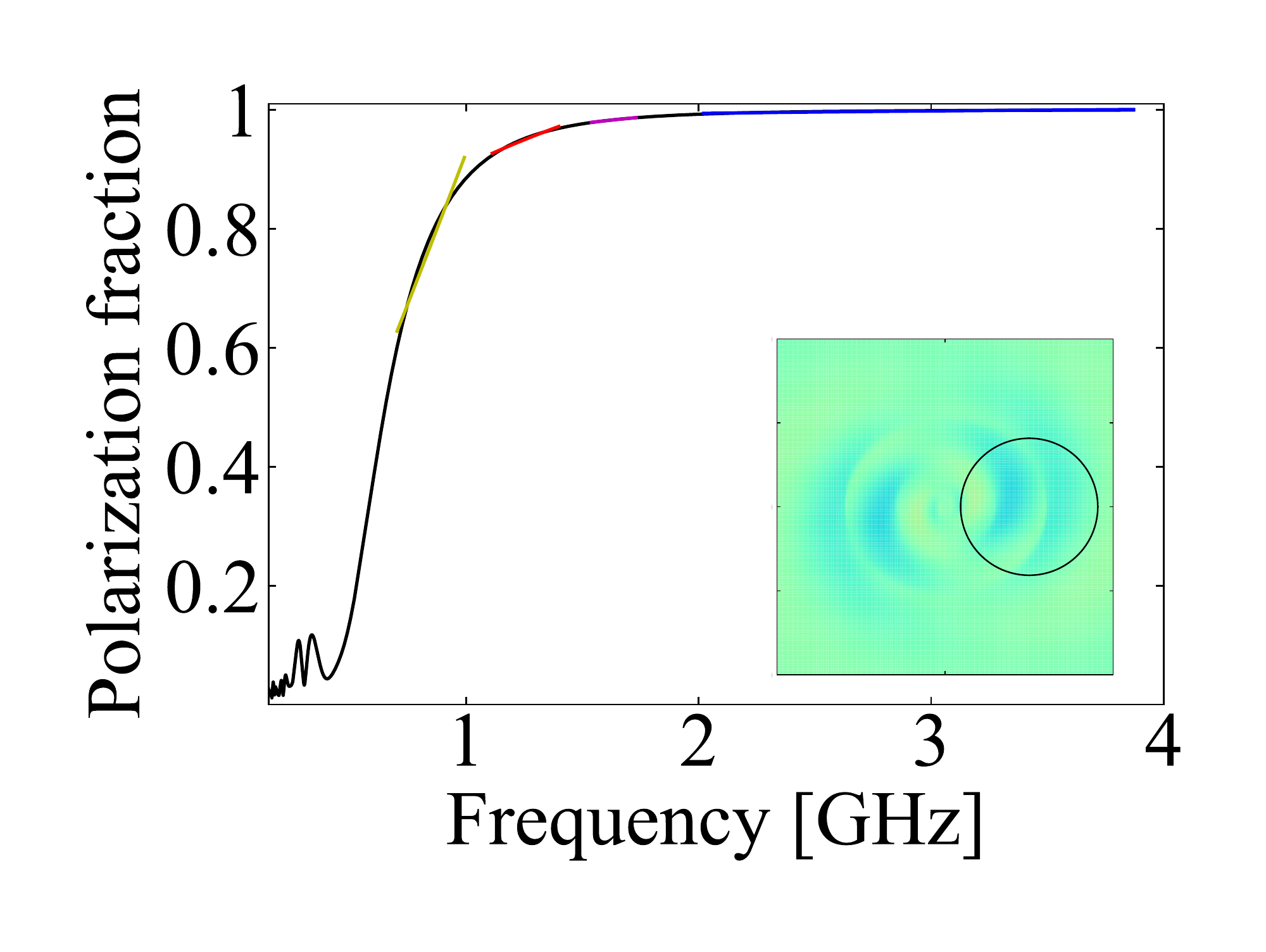}
          \hspace{0.5cm}
        \end{center}
      \end{minipage}\\    
    
      % 1_G
      \begin{minipage}{0.33\hsize}
        \begin{center}
         \includegraphics[width=1\textwidth,bb=0 0 576 433]{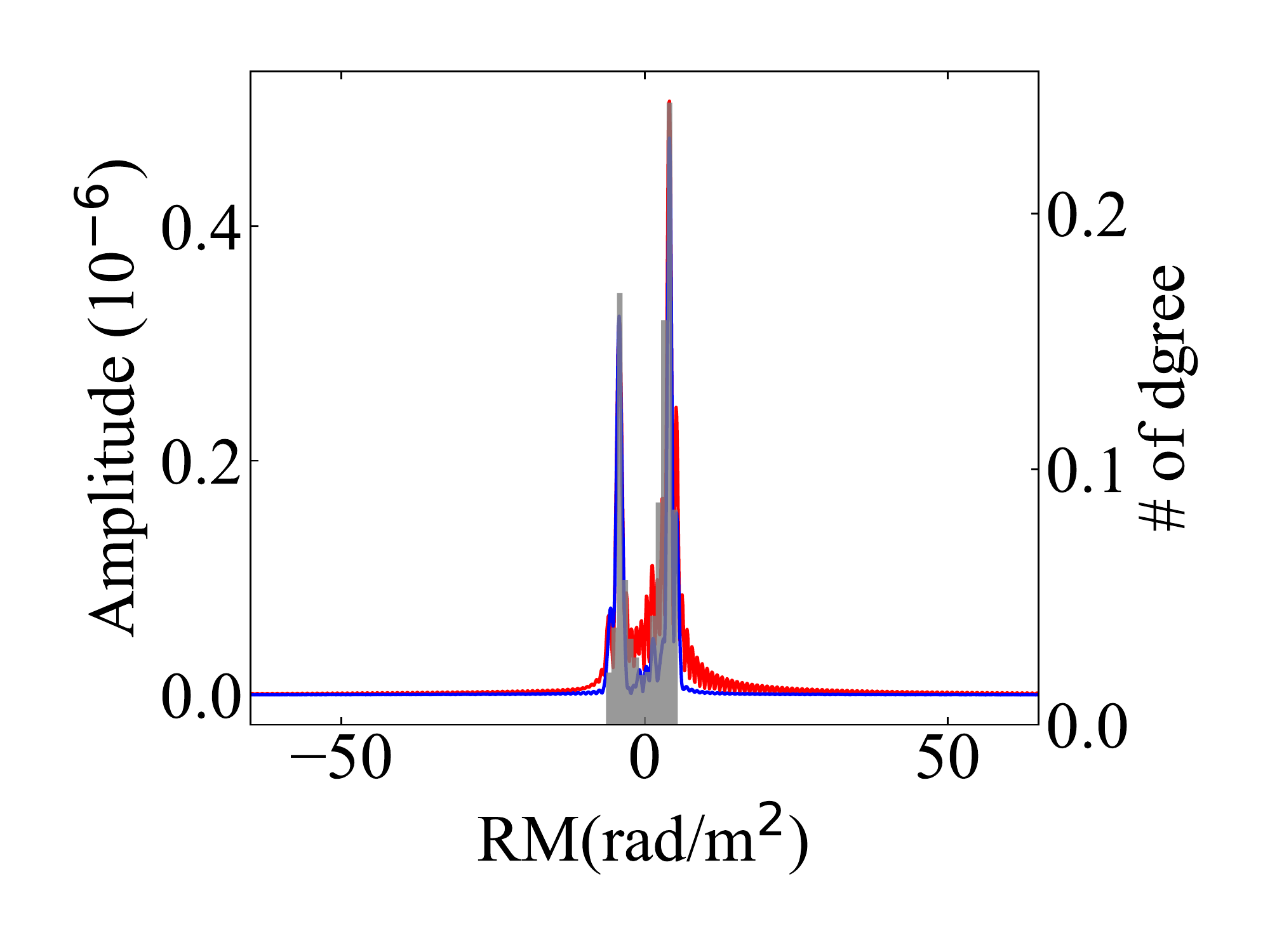}
        \end{center}
      \end{minipage}

      % 2_G
      \begin{minipage}{0.33\hsize}
        \begin{center}
          \includegraphics[width=1\textwidth,bb=0 0 576 432]{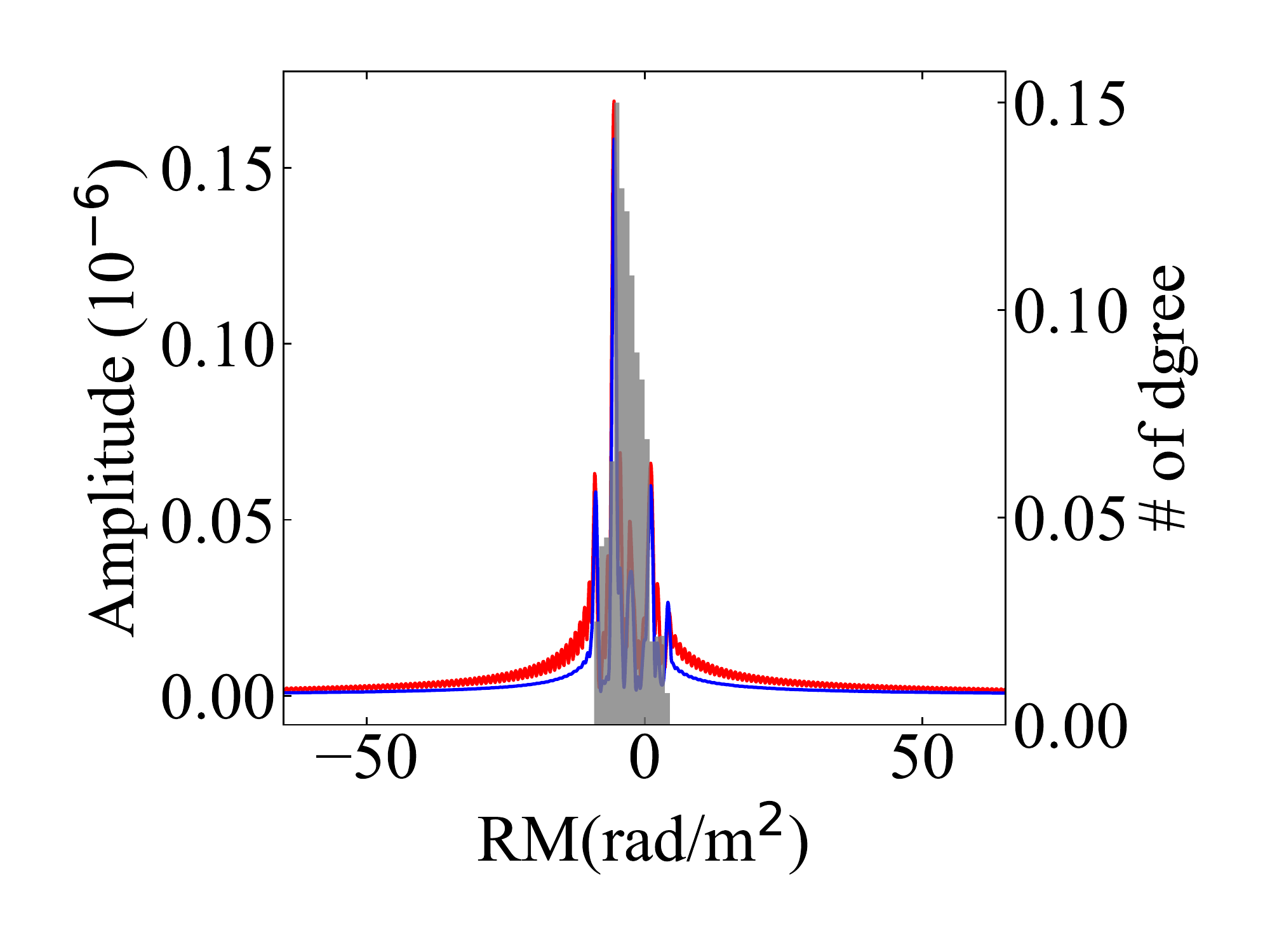}
        \end{center}
      \end{minipage}\\

      % 1_A
      \begin{minipage}{0.33\hsize}
        \begin{center}
         \includegraphics[width=1\textwidth,bb=0 0 576 433]{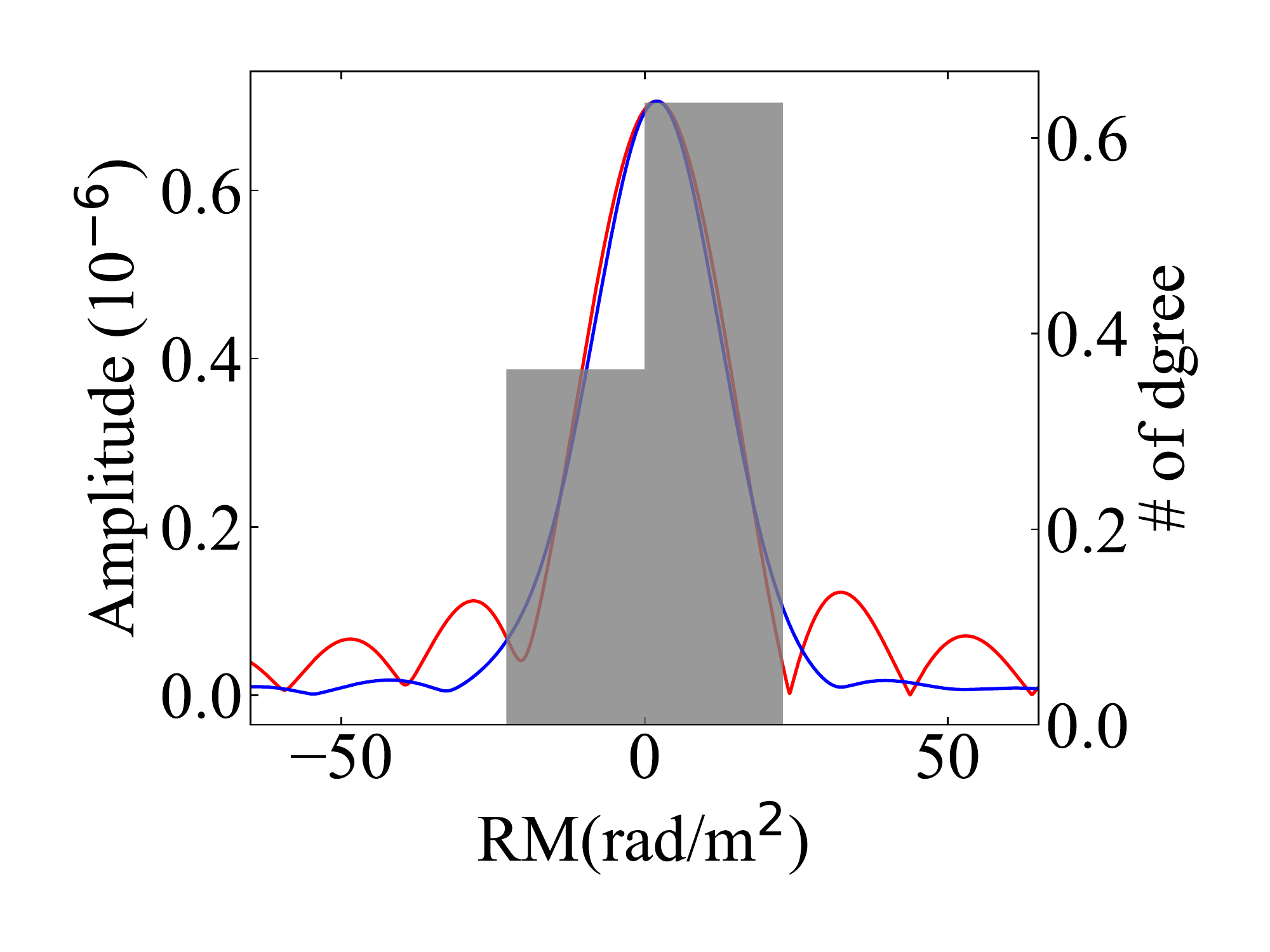}
         \hspace{0.5cm} ASS model
        \end{center}
      \end{minipage}

      % 2_A
      \begin{minipage}{0.33\hsize}
        \begin{center}
          \includegraphics[width=1\textwidth,bb=0 0 576 432]{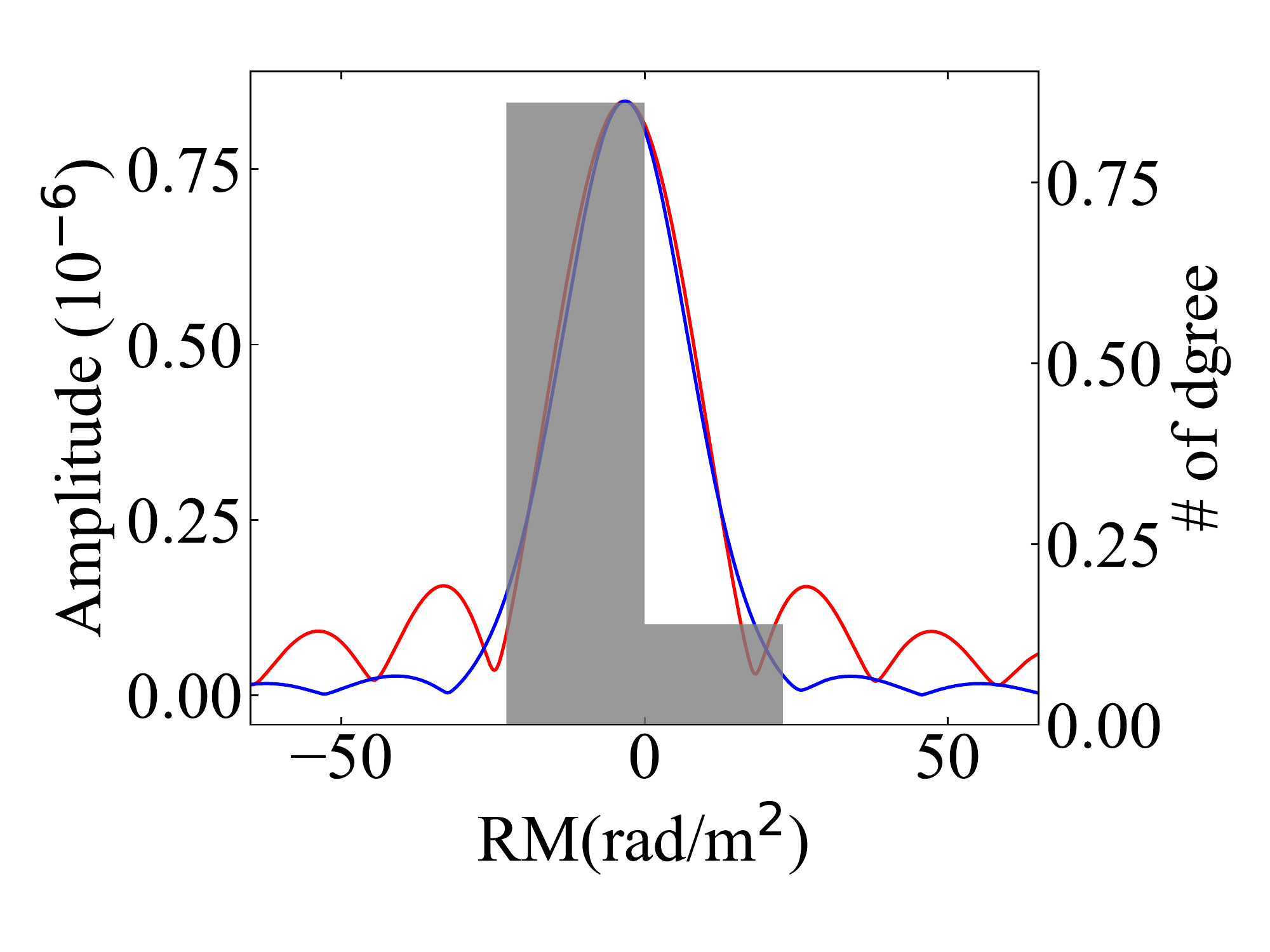}
          \hspace{0.5cm} BSS model
        \end{center}
      \end{minipage}\\
    \end{tabular}
    \end{center}
    \caption{Top panels: The polarization fraction as a function of the frequency. The left to right panels show the results for the inclination angles of 30$^\circ$ with ASS and BSS model, respectively, where the insets show the RM maps and the 1$^{\prime\prime}$ beam position is shown as the black circle. The color scale is the same as that in Figure \ref{fig:RM}. The both panels show the cases for $(z,x_1,y_1)=(1.0,5,0)$. The yellow, red, magenta and blue lines indicate the best fits that is the same as that in Figure \ref{fig:RMz1_x5y0_DP}.Center and bottom: Histograms of RM (gray) and simulated FDFs (red and blue) for different models of the DING and different frequency coverage. The gray color shows the histogram of RM, where the histograms are normalized by $1/(1+z)^2$. Red and blue lines show the dirty and cleaned FDFs, respectively. Panels from center to bottom show the results for the frequency coverage of low (150~MHz -- 700~MHz), mid (700~MHz -- 1800 MHz), respectively.}
    \label{fig:RM_AB_z1_x5y0_RMsys}
\end{figure*}

To assess the above speculation, we show the results for the ASS and BSS models (see \cite{2008A&A...477..573S}; the appendix of this paper). The results with the inclination angle of 30$^\circ$ and $(z,x_1,y_1)=(1.0,5,0)$ are shown in Figure \ref{fig:RM_AB_z1_x5y0_RMsys}. The top panels are the same as Figure \ref{fig:RMz1_x5y0_DP} but for ASS (left) and BSS (right) models. The middle and bottom panels are the same as Figure \ref{fig:RMz1_x5y0_RMsys} but for ASS (left) and BSS (right) models. The average and the standard deviation of RM within the beam are summarized in table~\ref{table_av_sd_model}.

As expected, the dispersion of RM caused by the positive and negative RM pairs within the beam for the ASS and BSS models are larger than that for the RING model (tables \ref{table_av_sd} and \ref{table_av_sd_model}), because the ASS and BSS models have more magnetic field reversals. A significant depolarization spectrum is seen as well (the top panels of Figure \ref{fig:RM_AB_z1_x5y0_RMsys}). As shown in the insets of the top panels, the ASS and BSS models possess positive and negative RM regions within the beam, so we expect multiple peaks in the FDF. Actually, the peaks can be resolved with the low-band data (the middle panels), and we can clearly see two peaks for the ASS model. Meanwhile, such multiple peaks are not apparent in the FDF with the mid-band data (the bottom panels), since the RM structure within the beam is smaller than the resolution of the FDF.

\begin{table}
\tbl{The average and the standard deviation of the observer-frame, in-beam RMs of the DING.}{%
\centering
  \begin{tabular}{ccccc} \hline
Model type & $(z,x_1,y_1)$ & $i$ & average & $\sigma_{\rm RM}$ \\
 & & (degree) & ( $\mathrm{rad/m^2}$) & ( $\mathrm{rad/m^2}$) \\
 \hline \hline
ASS &$(1.0,5,0)$ & 30 & 0.87  & 3.6 \\
BSS &$(1.0,5,0)$& 30 & -3.1 & 2.7 \\
\hline
BSS &$(0.1,0,8)$& 90 & 0.0  & 0.0 \\
BSS + D &$(0.1,0,8)$& 90 & 0.0 & 0.17 \\
BSS + Q &$(0.1,0,8)$& 90 & 0.0 & 0.17 \\
\hline
ASS &$(0.3,5,0)$& 30 & 3.6  & 9.4 \\
ASS + random &$(0.3,5,0)$& 30 & 3.5 & 10 \\
\hline
  \end{tabular}}\label{table_av_sd_model}
\end{table}

\subsubsection{Halo Fields}

We did not consider magnetic fields in the galactic halo. Previous works have shown that there are global magnetic fields in the halo of the Milky Way \citep{2010RAA....10.1287S, 2012ApJ...757...14J} and external galaxies \citep{2009IAUS..259..509H, 2009RMxAC..36...25K, 2013A&A...560A..42M}. Because of a lack of the halo component in this work, RM is exactly zero through the halo direction when we observe the DING in the edge-on view. This would be unlikely because the global electron and magnetic fields in the halo should have non-zero RM structures. Such RM can be small (say  $\lesssim 1~\mathrm{rad/m^2}$) because of an exponential decrease of the electron density toward the halo, supporting that the disk component is predominant for the DING.

\begin{figure*}
  \begin{center}
    \begin{tabular}{c}

      % 1_M
      \begin{minipage}{0.33\hsize}
        \begin{center}
         \includegraphics[width=1\textwidth,bb=0 0 576 433]{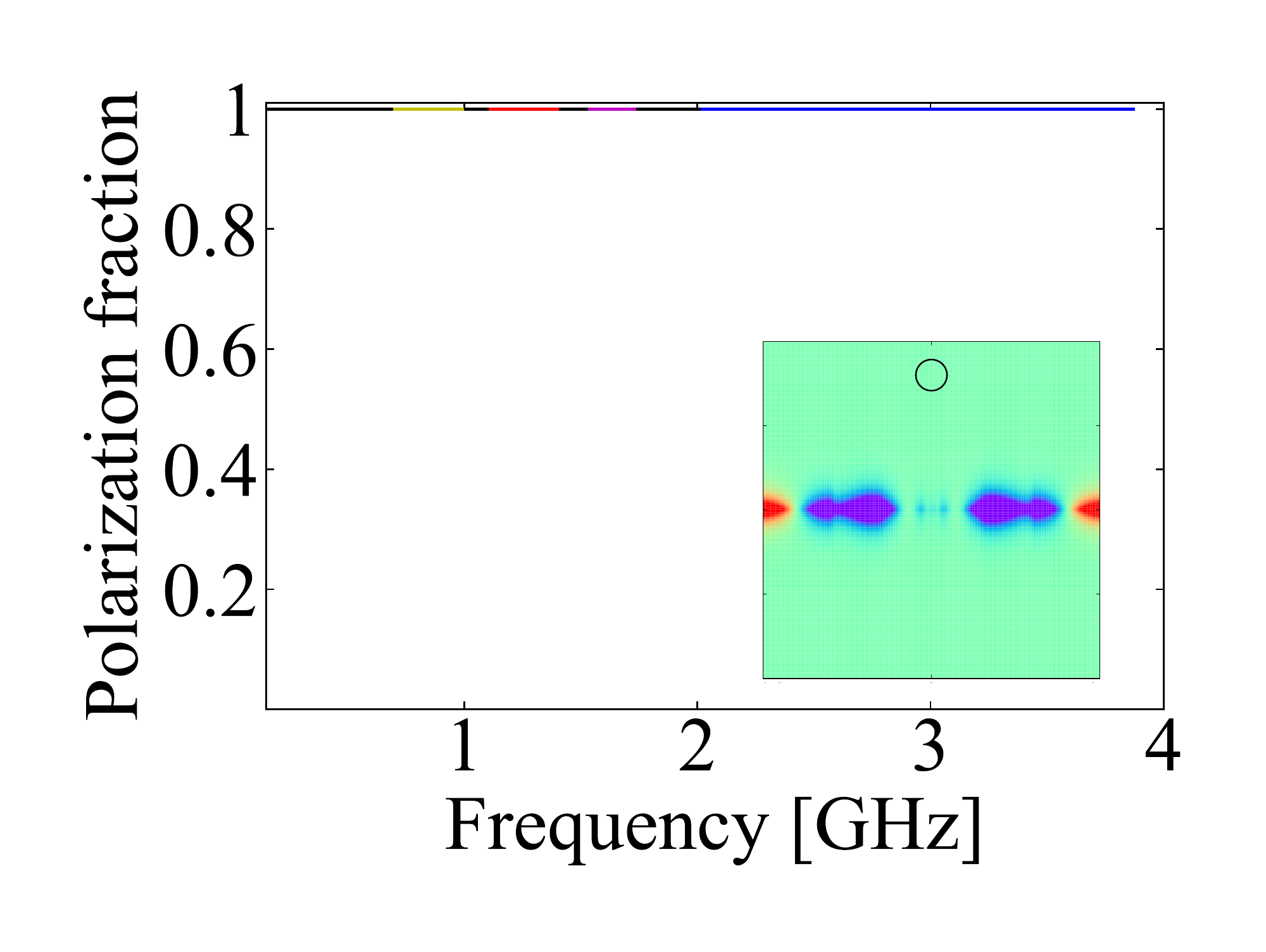}
        \end{center}
      \end{minipage}

      % 2_M
      \begin{minipage}{0.33\hsize}
        \begin{center}
          \includegraphics[width=1\textwidth,bb=0 0 576 432]{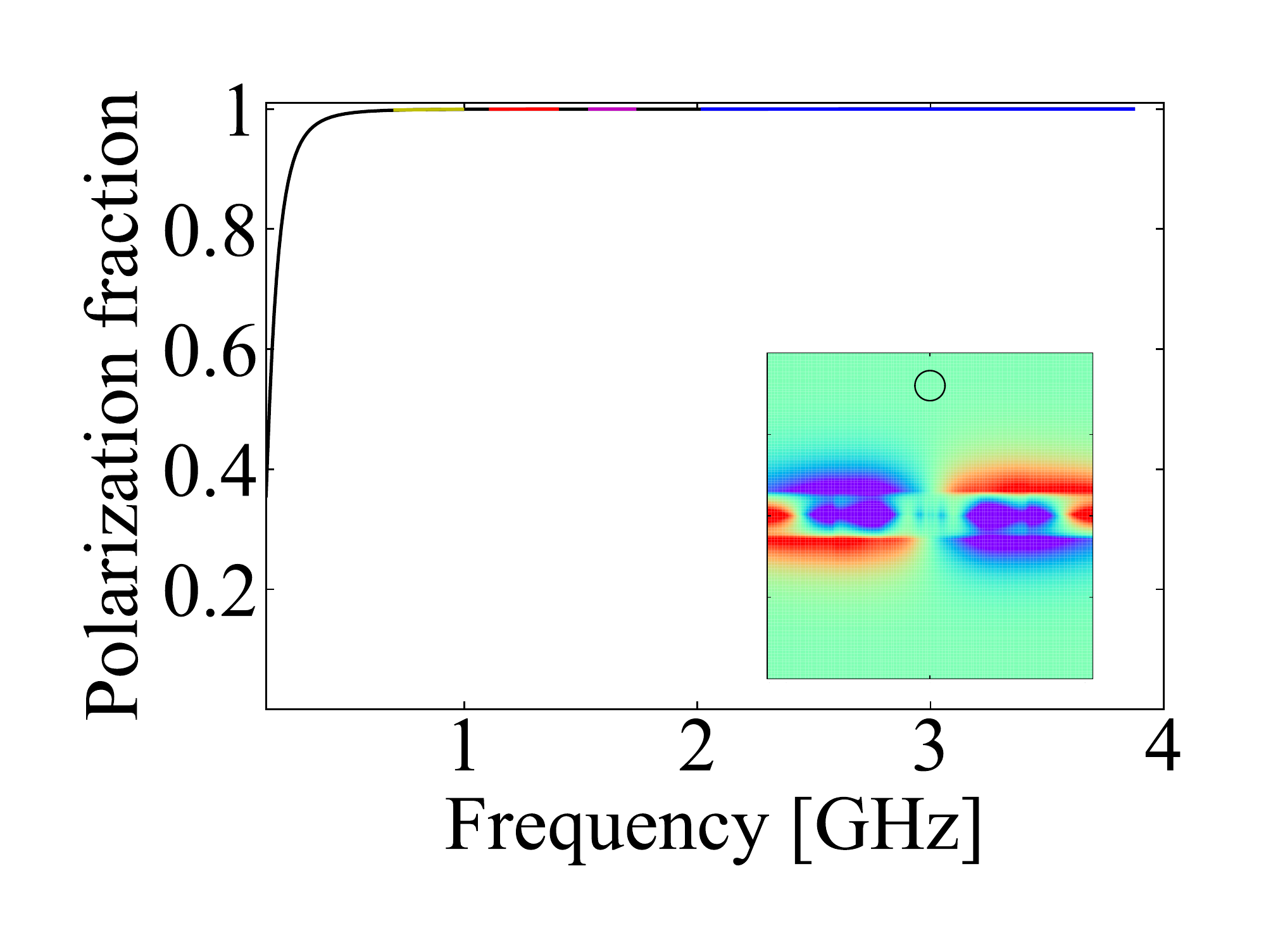}
        \end{center}
      \end{minipage}

      % 3_M
      \begin{minipage}{0.33\hsize}
        \begin{center}
          \includegraphics[width=1\textwidth,bb=0 0 576 432]{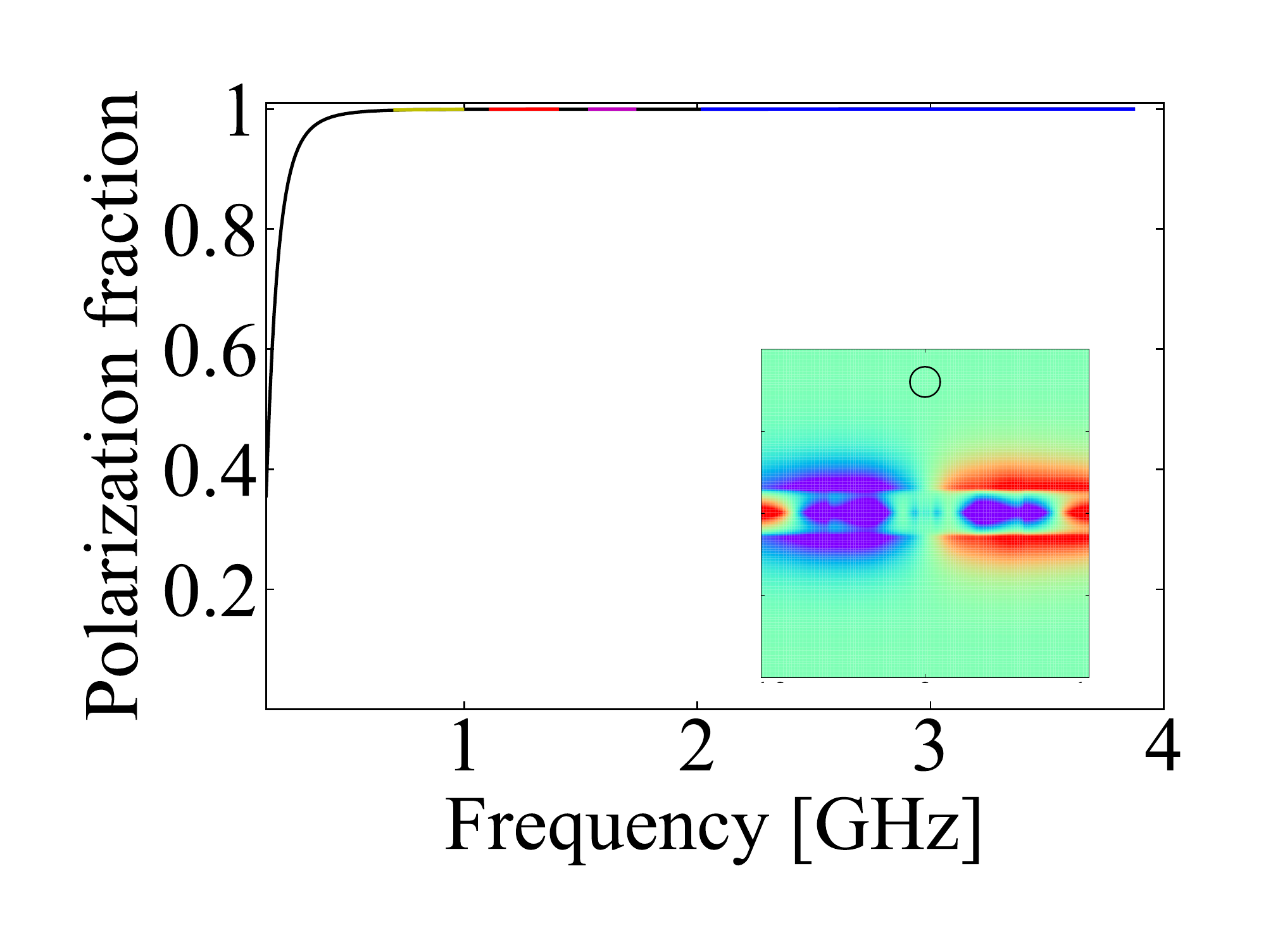}
        \end{center}
      \end{minipage}\\

      % 1_A
      \begin{minipage}{0.33\hsize}
        \begin{center}
         \includegraphics[width=1\textwidth,bb=0 0 576 433]{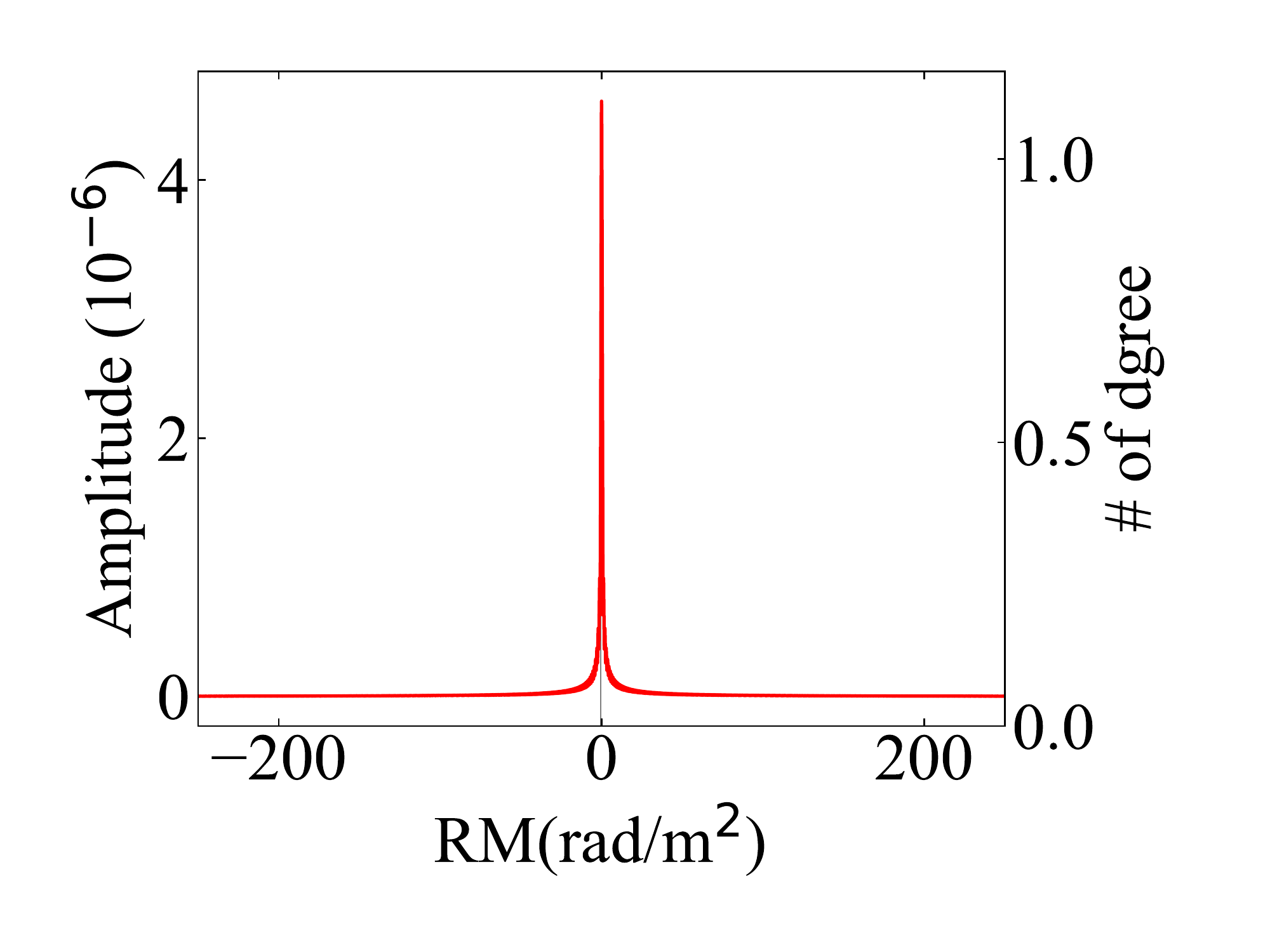}
         \hspace{1.6cm} BSS
        \end{center}
      \end{minipage}

      % 2_A
      \begin{minipage}{0.33\hsize}
        \begin{center}
          \includegraphics[width=1\textwidth,bb=0 0 576 432]{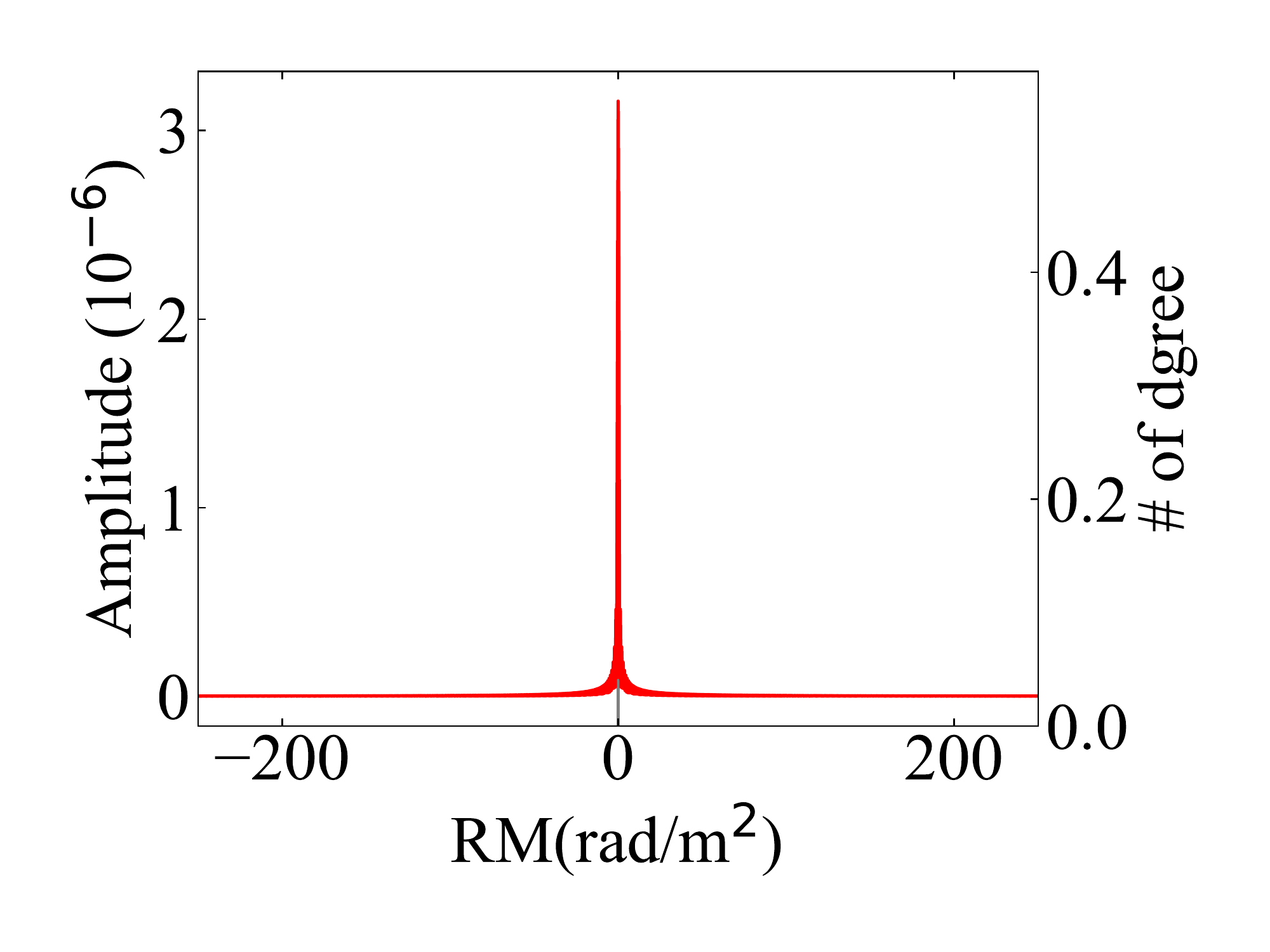}
          \hspace{1.6cm} BSS + D
        \end{center}
      \end{minipage}

      % 3_A
      \begin{minipage}{0.33\hsize}
        \begin{center}
          \includegraphics[width=1\textwidth,bb=0 0 576 432]{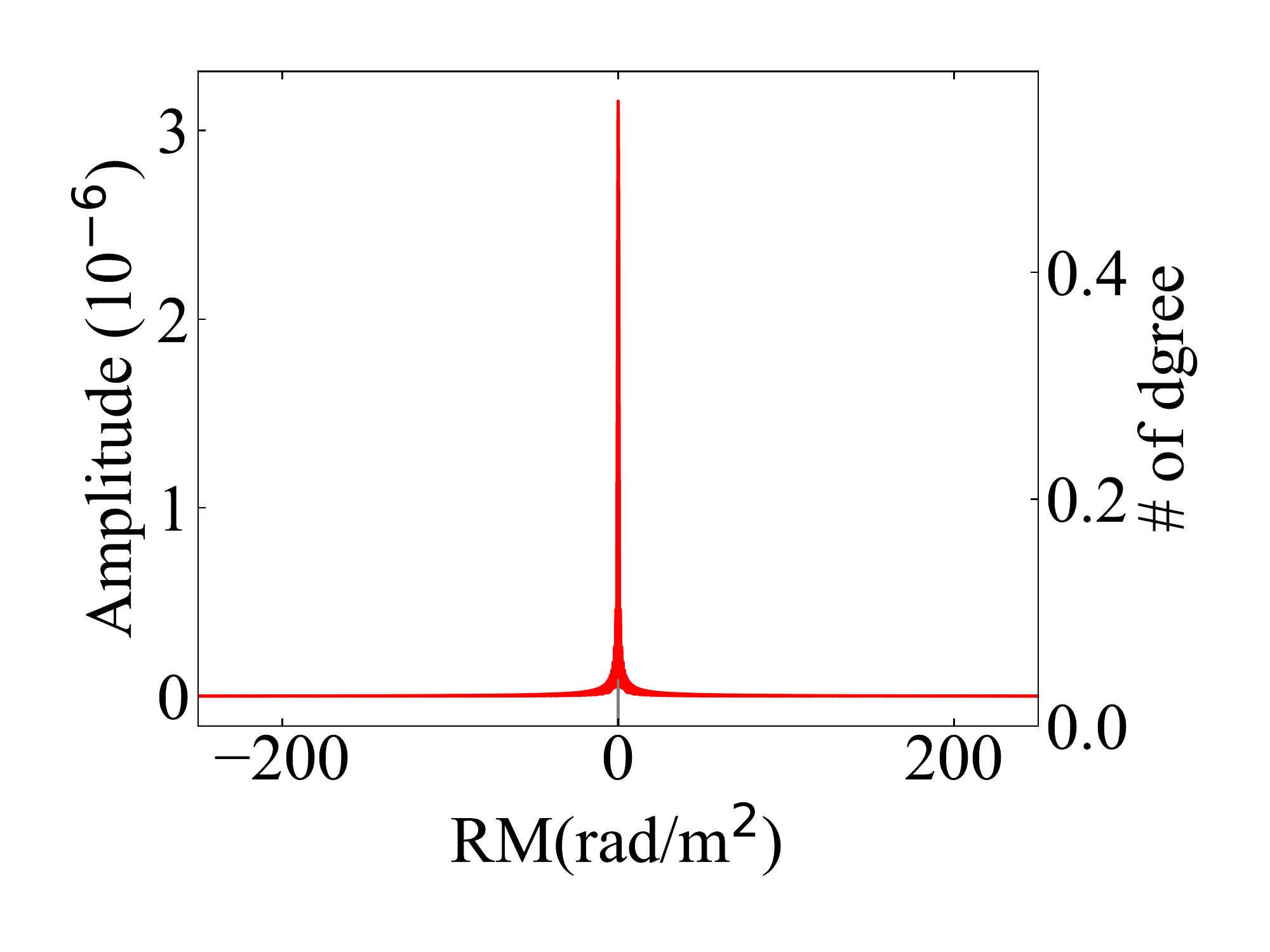}
          \hspace{1.6cm} BSS + Q
        \end{center}
      \end{minipage}
    \end{tabular}
    \end{center}
    \caption{Top: The polarization fraction as a function of the frequency. The left to right panels show the results for the inclination angles of 90$^\circ$ with BSS, BSS + D and BSS + Q model, respectively, where the insets show the RM maps and the 1$^{\prime\prime}$ beam position is shown as the black circle. The color scale is the same as that in Figure \ref{fig:RM}. The panels show the cases for $(z,x_1,y_1)=(0.1,0,8)$. The yellow, red, magenta and blue lines indicate the best fits that is the same as that in Figure \ref{fig:RMz1_x5y0_DP}. Bottom: Histograms of RM (gray) and simulated FDFs for different models of the DING (left to right). The gray color shows the histogram of RM, where the histograms are normalized by $1/(1+z)^2$. Red show the dirty.}
    \label{fig:RM_z01_x0y8_RMsys}
\end{figure*}

To examine the contribution of halo magnetic fields to the DING's depolarization quantitatively, we add a toroidal magnetic field with the asymmetric configuration in longitude and latitude relative to the galactic plane and the center, respectively (labelled D), or an axisymmetric magnetic field without reversals relative to the plane (labelled Q) \citep{2010RAA....10.1287S}. The results are shown in Figure \ref{fig:RM_z01_x0y8_RMsys}. The average and the standard deviation of RM within the beam are summarized in table~\ref{table_av_sd_model}.

This additional halo component produces more complex RM structures in the FDF in general, regardless of the model (not shown). A clear difference made by the halo field is seen when we view the DING through the edge-on. The top panels in Figure \ref{fig:RM_z01_x0y8_RMsys} show the frequency dependence of the polarization fraction for the inclination angles of 90$^\circ$ and $(z,x_1,y_1)=(0.1,0,8)$. The left to right panels compare the results with BSS, BSS + D and BSS + Q models, respectively, showing that the additional halo component causes depolarization at low frequency because the BSS + D and BSS + Q models have non-zero standard deviations within the beam (table~\ref{table_av_sd_model}). The bottom panels in Figure \ref{fig:RM_z01_x0y8_RMsys} show the Faraday spectra respectively corresponding to the top panels. Because the RM structure within the beam is finer than the resolution of the Faraday depth, the FDF apparently indicates one component. This demonstrates that, even when the effect of DING is not clearly seen in the Faraday spectrum, we may be able to find depolarization caused by a weak magnetic field such as a halo magnetic field at very low frequencies.

Note that galaxies including DINGs probably possess the circumgalactic medium (CGM) that extends to 400 -- 500 kpc from the center \citep{2017ARA&A..55..389T, 2020MNRAS.498.3125P}. Although the typical values of the density and magnetic field of the CGM are not clear, CGM's RM could be small like the gas in the halo. With realistic halo gas and RM, our results relating to the filling factor will be improved.

\subsubsection{Turbulence}

We did not consider small-scale random and turbulent magnetic fields which have been addressed in the previous works as a source of depolarization (e.g., \cite{2008A&A...477..573S}). If we consider turbulent magnetic fields, our results will largely change. More specifically, our results show the minimum case and an additional depolarization takes place by the turbulent component. For example, for the case with $i=30^{\circ}$ in Figure \ref{fig:RMz1_x5y0_DP}, the polarization fraction remains almost unity at above 1000 MHz. This is because the standard deviation of intrinsic RM within the beam, 5.6 $\mathrm{rad/m^2}$, does not significantly rotate the polarization angles at those frequencies. Such a result varies if there is an additional RM dispersion induced by turbulent magnetic fields. The significance of this addition depends on the model of turbulence and hence accurate modeling of turbulent magnetic fields is crucial \citep{2013ApJ...767..150A}. 

To explore the contribution of the random component, we adopt a simple model that the random component is homogeneous and follow a Gaussian distribution in strength with an average of zero and a scatter of $\sigma_b/\sqrt{3}$ in each dimension, where $\sigma_b$ is the standard deviation of the random component. We set the parameter, $\sigma_b = 3$ ($\mu$G), based on the previous work \citep{2008A&A...477..573S}.

\begin{figure*}
  \begin{center}
    \begin{tabular}{c}
    
      % 2_M
      \begin{minipage}{0.33\hsize}
        \begin{center}
          \includegraphics[width=1\textwidth,bb=0 0 576 432]{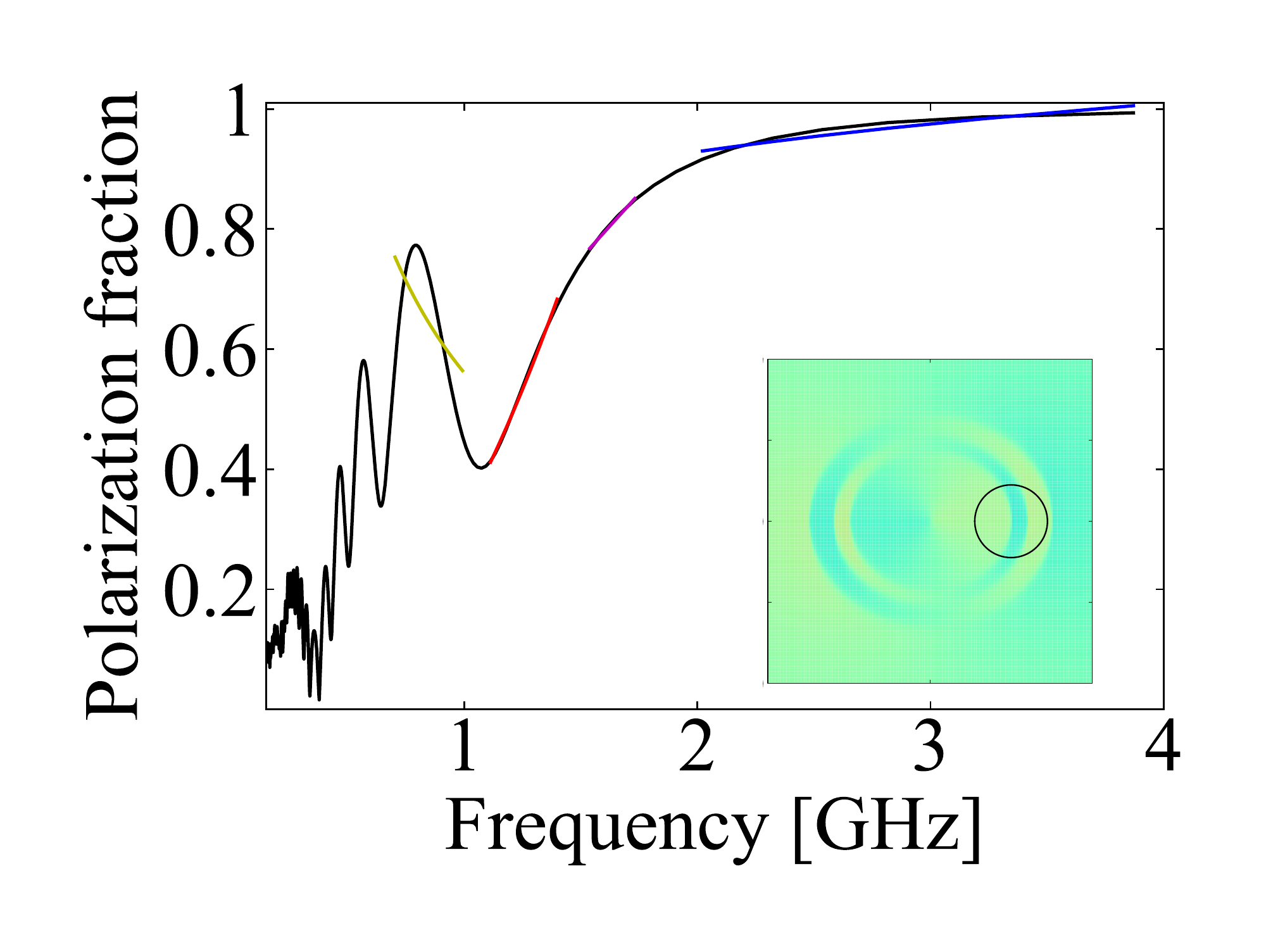}
        \end{center}
      \end{minipage}

      % 3_M
      \begin{minipage}{0.33\hsize}
        \begin{center}
          \includegraphics[width=1\textwidth,bb=0 0 576 432]{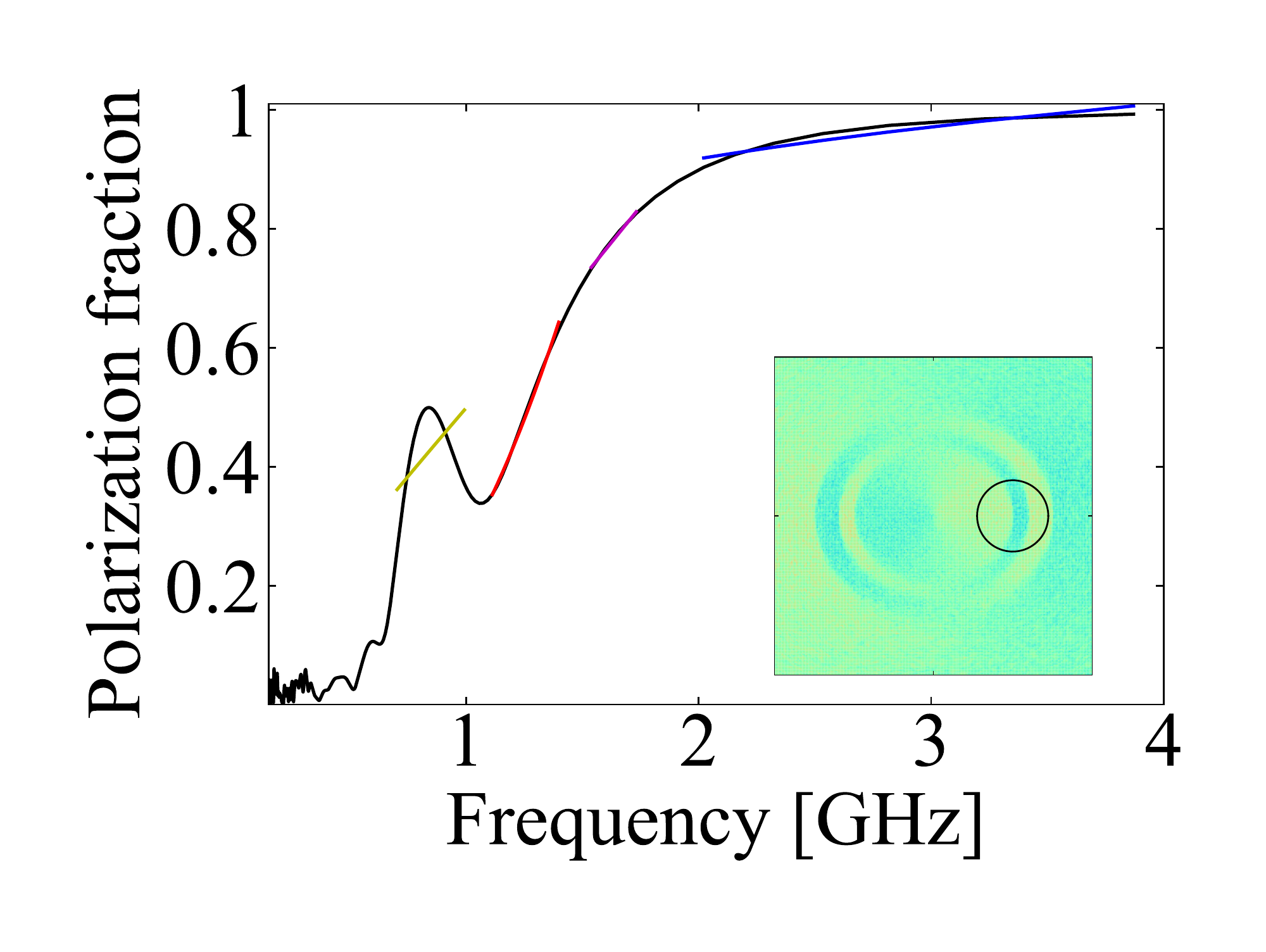}
        \end{center}
      \end{minipage}\\
      
      % 1_G
      \begin{minipage}{0.33\hsize}
        \begin{center}
         \includegraphics[width=1\textwidth,bb=0 0 576 433]{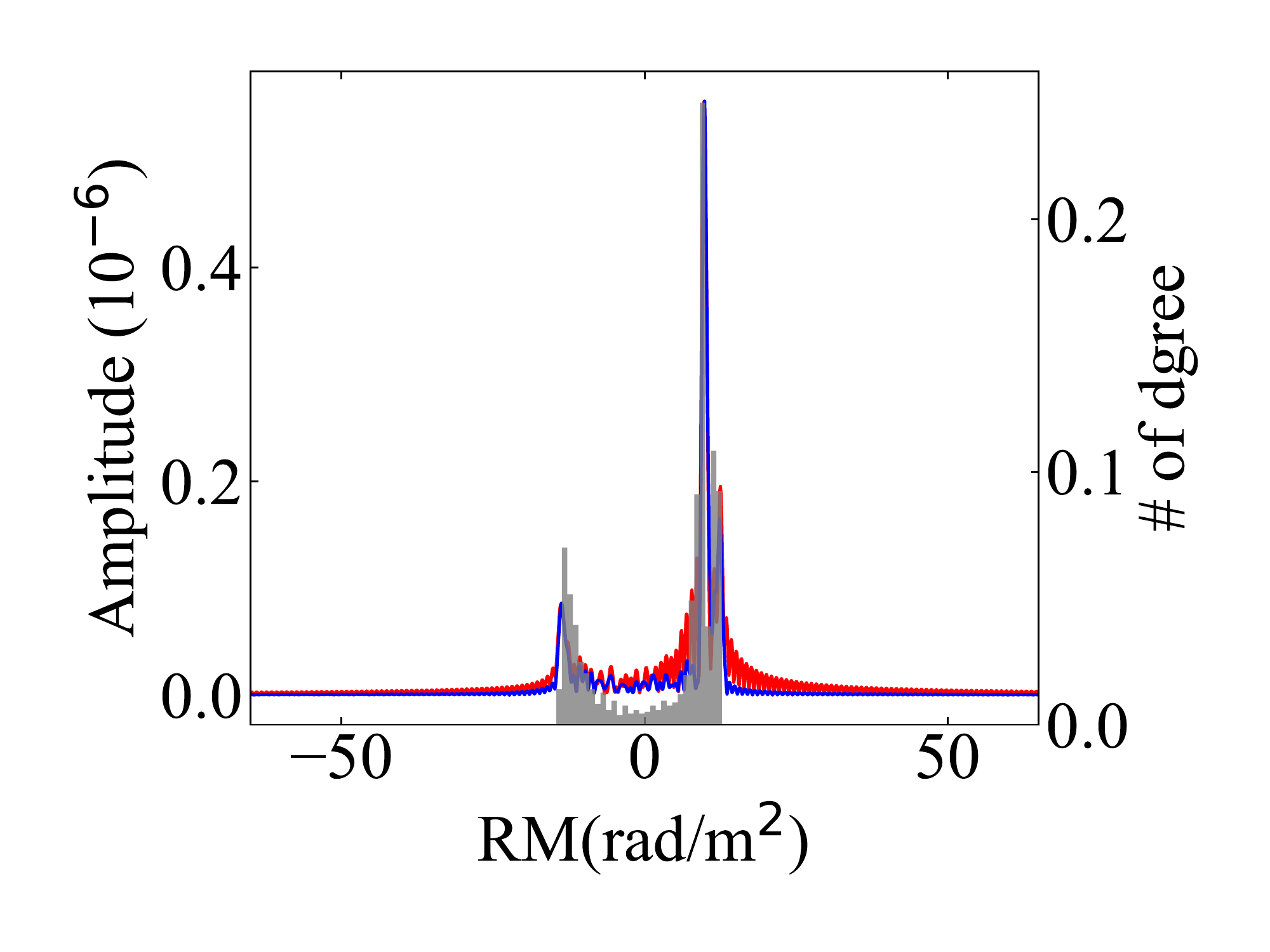}
        \end{center}
      \end{minipage}

      % 2_G
      \begin{minipage}{0.33\hsize}
        \begin{center}
          \includegraphics[width=1\textwidth,bb=0 0 576 432]{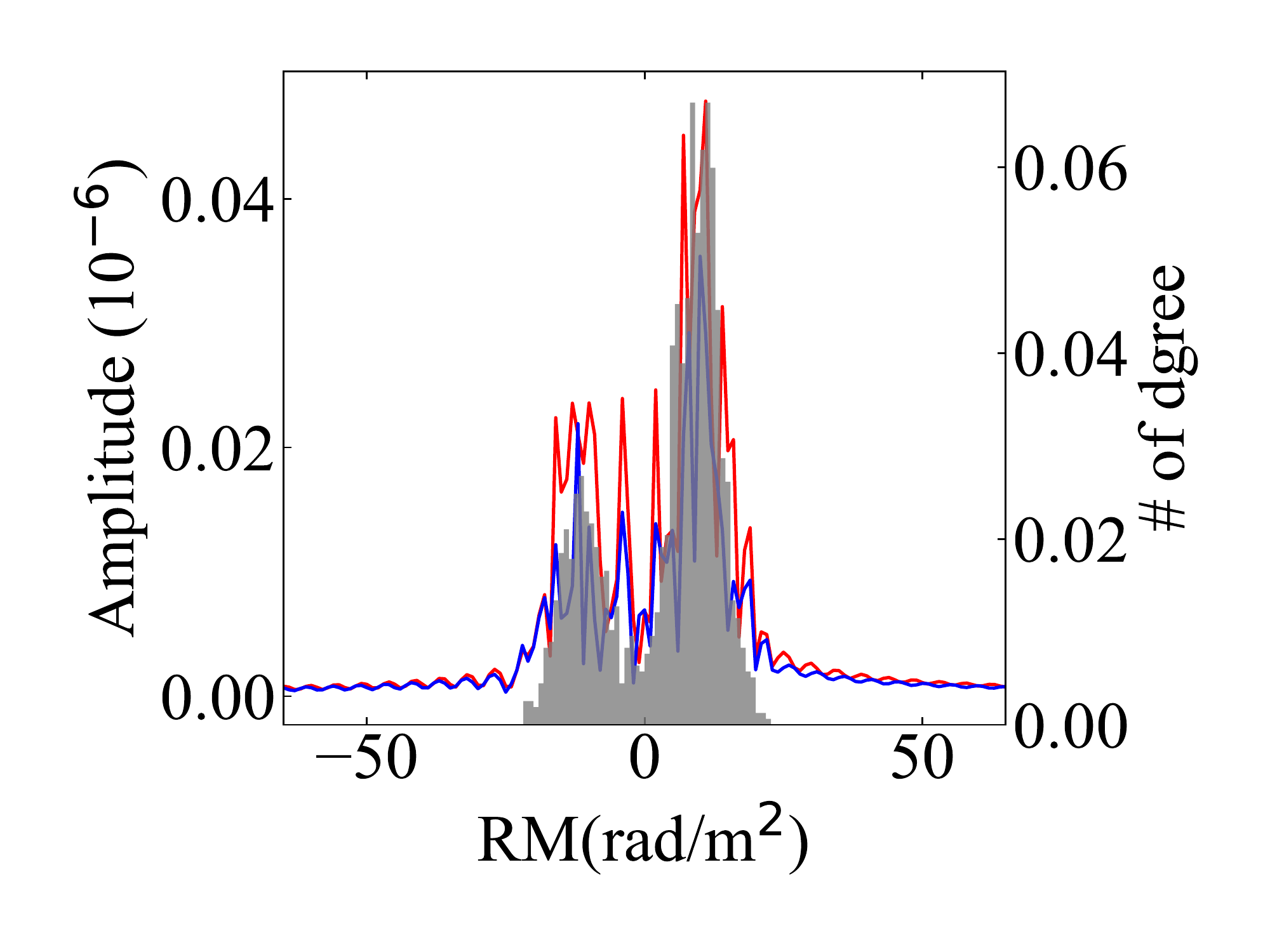}
        \end{center}
      \end{minipage}\\

      % 1_A
      \begin{minipage}{0.33\hsize}
        \begin{center}
         \includegraphics[width=1\textwidth,bb=0 0 576 433]{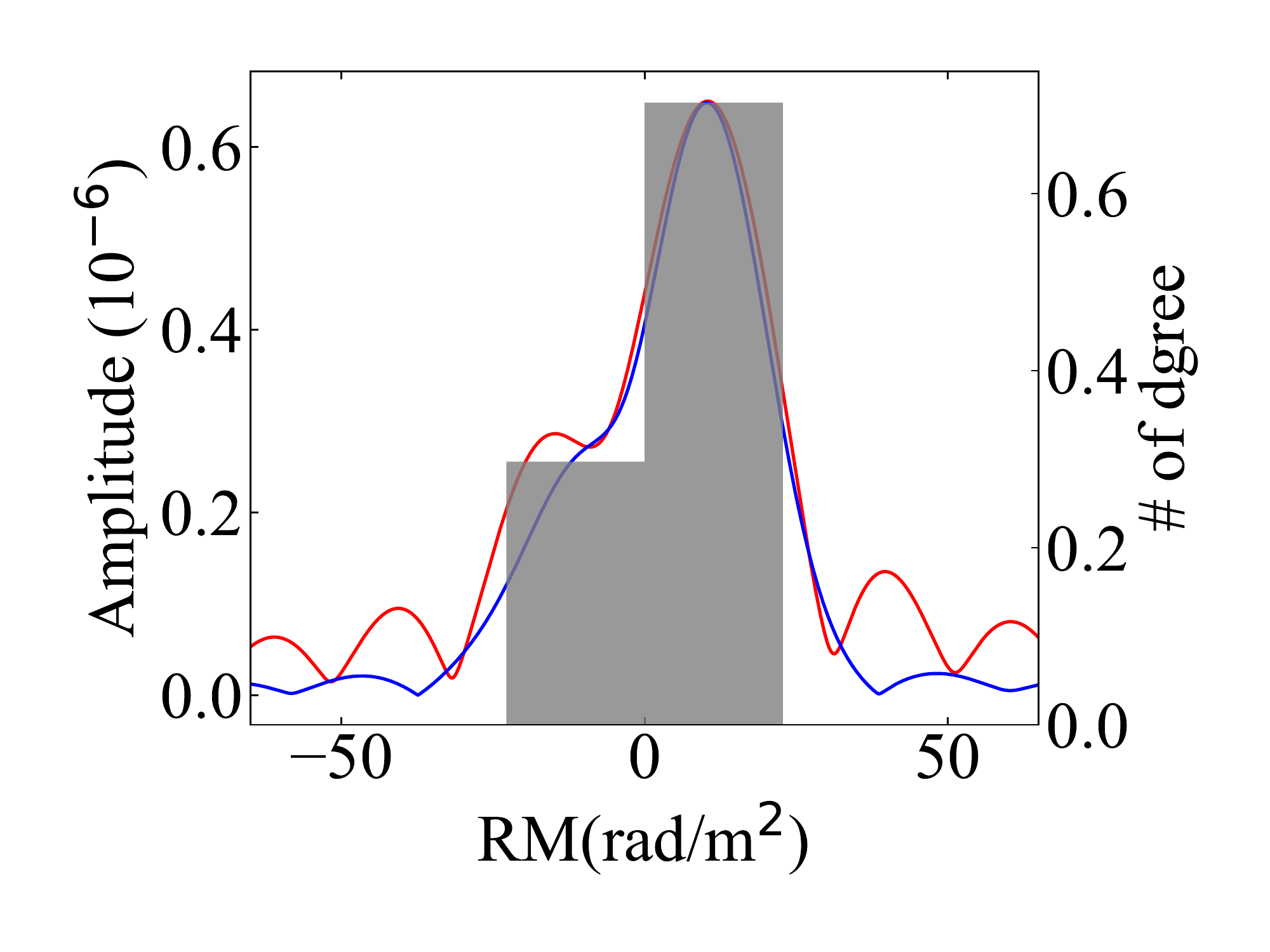}
         \hspace{0.5cm} ASS model
        \end{center}
      \end{minipage}

      % 2_A
      \begin{minipage}{0.33\hsize}
        \begin{center}
          \includegraphics[width=1\textwidth,bb=0 0 576 432]{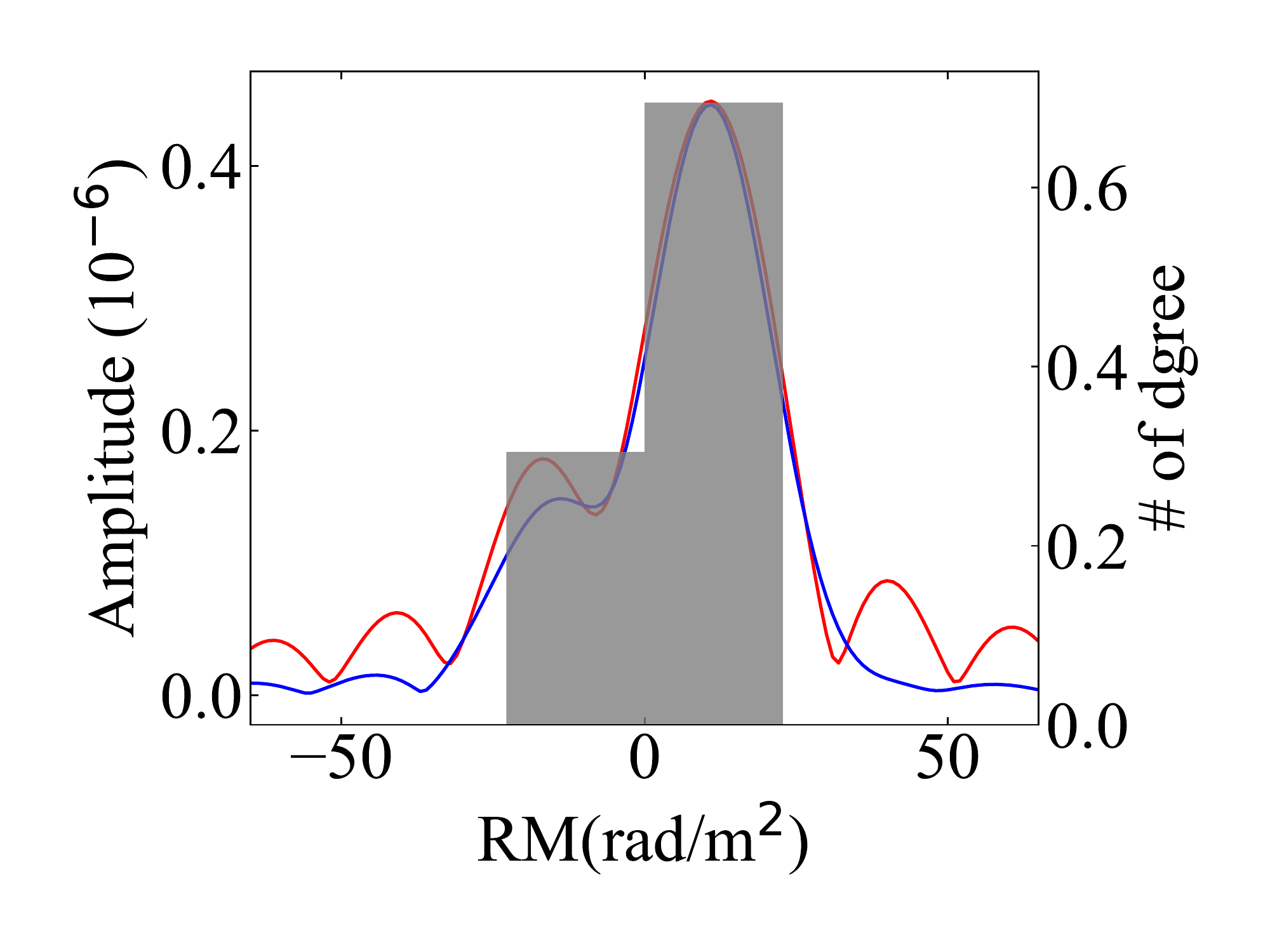}
          \hspace{0.5cm} ASS + random model
        \end{center}
      \end{minipage}\\
    \end{tabular}
    \end{center}
    \caption{Top: The polarization fraction as a function of the frequency. The left to right panels show the results for the inclination angles of 60$^\circ$ with, ASS and ASS + random model, respectively, where the insets show the RM maps and the 1$^{\prime\prime}$ beam position is shown as the black circle. The color scale is the same as that in Figure \ref{fig:RM}. The panels show the cases for $(z,x_1,y_1)=(0.3,5,0)$. The yellow, red, magenta and blue lines indicate the best fits that is the same as that in Figure \ref{fig:RMz1_x5y0_DP}. Bottom: Histograms of RM (gray) and simulated FDFs for different models of the DING (left to right). The gray color shows the histogram of RM, where the histograms are normalized by $1/(1+z)^2$. Red and blue lines show the dirty and cleaned FDFs, respectively.}
    \label{fig:RM_AR_z1_x5y0_RMsys}
\end{figure*}

The results for the ASS model and the ASS + random field model are compared to each other in Figure \ref{fig:RM_AR_z1_x5y0_RMsys}. The average and the standard deviation of RM within the beam are summarized in table~\ref{table_av_sd_model}. As expected, the random component induces more depolarization, because the random component increases the standard deviation of RM within the beam. The FDF has two peaks for both the ASS and ASS + random models, but the latter indicates a broader FDF with more spikes. The random magnetic field can smooth the coherent RM structure formed by the global magnetic field, because random components have a Gaussian distribution. Thus, the amplitude of the FDF will become lower and the width of the FDF will become wider, but remaining the split of FDF. We note that we see one Gaussian RM structure if RM structure has no global component.

\subsection{Quantification of Depolarization}

The polarization spectral index $\beta$ of $p \propto \nu^\beta$ has been adopted to quantify the degree of depolarization (e.g., \cite{2014ApJ...795...63F}). We derived the best-fit values of $\beta$ for the four frequency bands (color lines in Figure \ref{fig:RMz1_x5y0_DP}), but it demonstrates that the linear fit is unreasonable to represent the oscillation of the polarization fraction caused by the DING. Such oscillation happens when the DING's RM is large. Here, as already shown, a larger inclination angle causes a larger RM, because not only the path length through the disk but also its $B_{\parallel}$ component becomes larger. Hence, a larger inclination angle results in stronger depolarization, because the standard deviation of intrinsic RM for the DING becomes larger. However, a larger inclination angle can reduce the chance for the beam to overlap the disk (see Figure~\ref{fig:Depo}), i.e. reduce the filling factor from unity. As a result, two components of polarization are found within a beam; one is the emission which is significantly depolarized by passing through the galactic plane with a large RM, showing a decaying oscillation in the profile of the polarization fraction. The other is the emission that can reach us without passing through the galactic plane, determining the convergence of the observed polarization fraction at the value of 1 - the filling factor.

The “Burn law" \citep{1966MNRAS.133...67B} is another analytic formula to quantify the depolarization. The formula is written as
\begin{equation}
\label{Burnlaw}
p(\lambda) = p_0(\lambda)\exp{(-2\sigma^2\lambda^4)},
\end{equation}
where $\lambda$ is the wavelength and $\sigma$ is the standard deviation of RMs within a considering beam. Obviously, the exponential formula is difficult to reproduce the non-monotonic change of the polarization fraction caused by the DING. But this discrepancy is reasonable because RMs within the beam do not follow the Gaussian distribution which is assumed in the Burns law. Moreover, a condition that a typical size of incoherent structures is sufficiently smaller than the beam is not satisfied.

These results indicate that the above traditional methods can only quantify a part of the DING's broadband depolarization properties at relatively high frequency of $\sim$GHz. Careful examination of the profile of the polarization fraction at a few to several hundred MHz is important to understand the depolarization caused by the DING. Fitting of the profile of the polarization fraction with advanced models such as the two components model (see Appendix of \cite{2014ApJ...795...63F}) would be useful to extract the features caused by the filling factor.

\subsection{Shapes of the Faraday Spectrum}\label{FDFdiscuss}

The shape of the FDF obtained from RM CLEAN is similar to the shape of the histogram of RM within the beam. While the histogram is the normalized histogram of RMs for the pixels within the beam, the FDF is the sum of the FDFs for the pixels within the beam. Because we consider a uniform linear polarization within the beam, the sum of the FDFs results in the histogram of Faraday depths for the pixels within the beam and the histogram can be compared to the normalized histogram of RMs. In reality, the shape of the FDF can depart from the shape of the histogram, if the amplitude of linear polarization is very inhomogeneous within the beam. This situation can occur when the emission has a significant structure, such as radio lobes, or when there is strong intrinsic depolarization due to small-scale turbulence. Another factor that causes the departure is numerical artifacts in Fourier transform. We can see the artefact at the edge of the histogram in the top-middle panel of Figure \ref{fig:RMz1_x5y0_RMsys}. We speculate that the peak is spurious that appears at the edge when we reproduce a square wave using a sine wave in the Fourier series expansion. A sophisticated method of Faraday tomography is required to make sure that we obtain the FDF which accurately reproduces the shape of the histogram.

We find that the beam-averaged RM of the DING deviates from ${\rm RM_{peak}}$, Faraday depth at the peak of the FDF. The reason why the peak of the FDF stands at a non-average RM position within the beam is that the region through which polarized emission passes contains many $0\ \mathrm{rad/m^2}$ regions. The discrepancy is, however, not surprised because the mean value and the mode value of the distribution are not always the same for a statistical distribution like the histogram. In the future, we need to perform Faraday RM synthesis for millions of sources obtained from large surveys. A choice of ${\rm RM_{peak}}$ as the RM toward the source is reasonable because that is easy to obtain from a simple Faraday RM synthesis (and RM CLEAN as an option). But it is not reasonable to refer ${\rm RM_{peak}}$ for the study of magnetic fields of DINGs because ${\rm RM_{peak}}$ (the mode value) can be far from the average RM (the mean value, i.e. the first moment of the FDF) which is a more representative value for the study of DING's magnetic fields. Another problem relating to the RM catalog is that we can obtain multiple FDF peaks (Figure \ref{fig:RMz1_x5y0_RMsys}) depending on the filling factor, although we consider a single background source. The split of a single source into multiple sources by a DING impacts on the studies of, for example, modeling of the foreground Milky Way RM map, study of the luminosity distribution function, and exploration of the intergalactic magnetic field (IGMF). Examining the profile of the polarization fraction as well as referring to optical absorption lines are promising ways to identify the existence of DINGs along the line of sight.

\subsection{Redshift Dependence of $\sigma_{\rm RM}$}\label{staticdis}

\begin{figure}
  \begin{center}
  	\includegraphics[width=1\linewidth,bb=0 0 576 432]{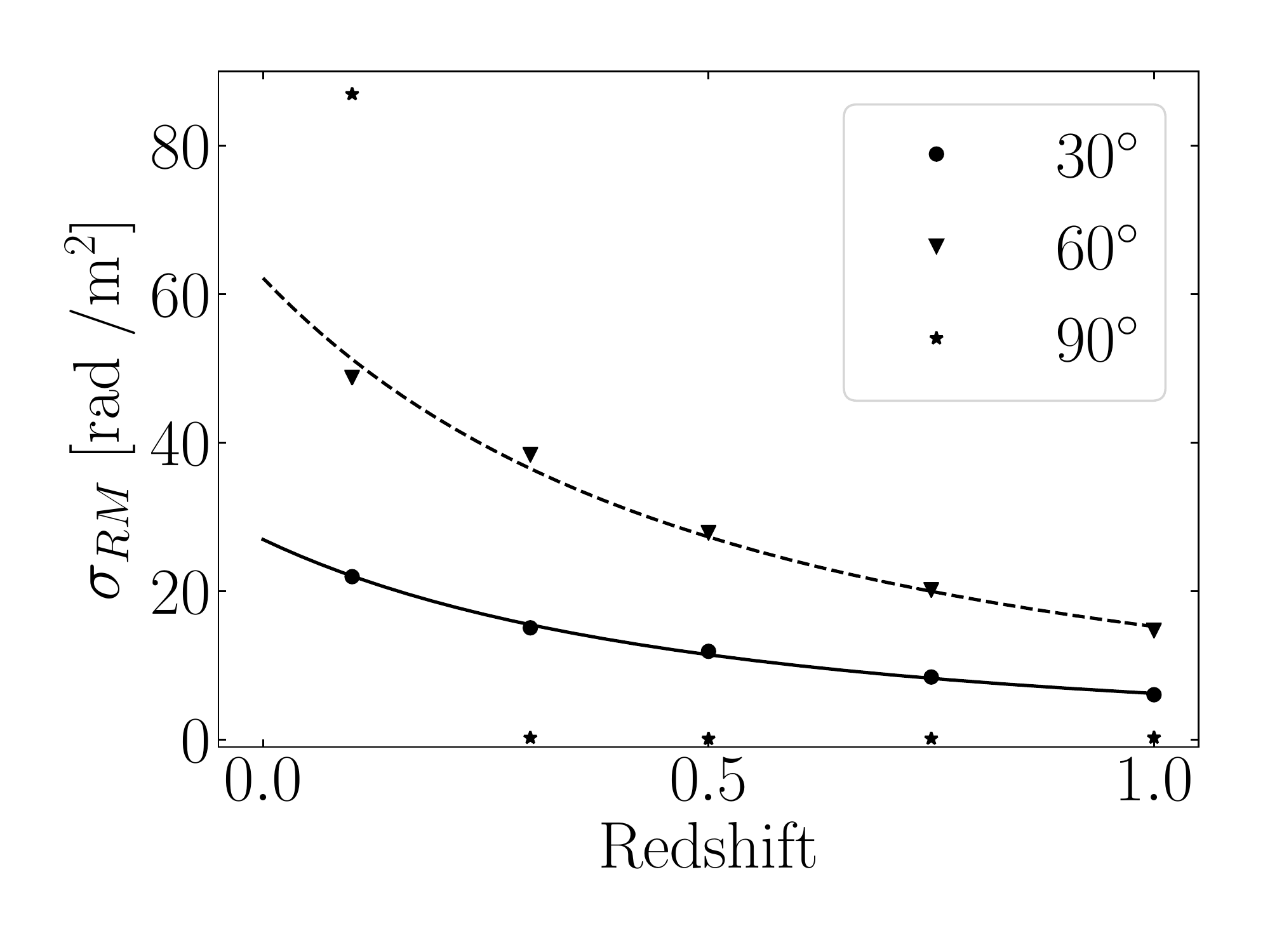}
  \end{center}
  \caption{Same as the control run in Figure \ref{fig:MCRMsig_fit} but separately shows the results of a different inclination angle, the circles for $i=30^\circ$, the triangles for $i=60^\circ$, and the stars for $i=90^\circ$. The solid and dashed lines indicate the best-fit power laws for $i=30^\circ$ and $60^\circ$, respectively.
  }
\label{fig:MCRM_inc}
\end{figure}

The standard deviation of DING's RM, $\sigma_{\rm RM}$, depends on the redshift (Section \ref{statis}). But the dependence, $k\sim 3$ for $\sigma_{\rm RM} \propto (1+z)^{-k}$, deviates from the expectation of $k\sim 2$ based on the wavelength-squared dependence of the RM. We claim that this steeper redshift dependence relates to the change of the physical size of the beam, i.e. the filling factor of the DING in the beam. To understand the effect of the filling factor more clearly, we again perform the control runs in which we fix the physical size of the beam. Since the filling factor significantly depends on the inclination angle of the DING, we separately show the results of the Monte-Carlo simulations for different inclination angles. The results are shown in Figure \ref{fig:MCRM_inc}. As expected, we obtain  $k \sim 2.1$ and $k \sim 2.0$, for  $i=30^{\circ}$ and $i=60^{\circ}$, respectively, both in good agreement with $k\sim 2$ based on the wavelength-squared dependence. Meanwhile, for $i=90^{\circ}$with a small filling factor, we find that the result does not follow the expected dependence; $\sigma_{\rm RM}$ of $\sim 80~\mathrm{rad/m^2}$quickly decreases as the redshift increases and then it reaches about 0 $\mathrm{rad/m^2}$at $z\gtrsim 0.3$. 

\begin{figure}
  \begin{center}
  	\includegraphics[width=1\linewidth,bb=0 0 576 432]{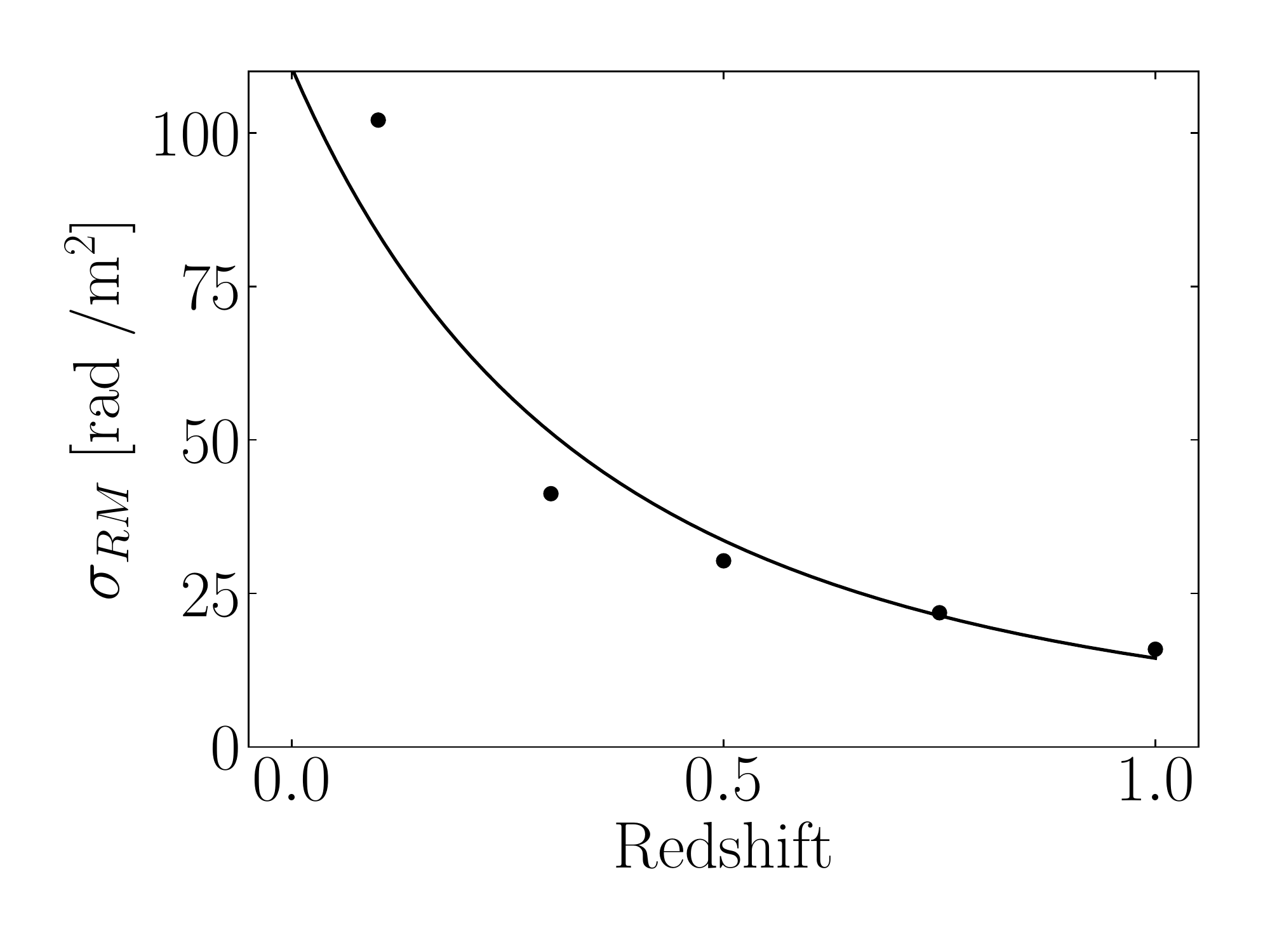}
  \end{center}
  \caption{Same as the control run in Figure \ref{fig:MCRMsig_fit} but the sum of only the cases for  $i=30^\circ$, $60^\circ$, and $90^\circ$. The solid line indicates the best-fit power laws.
  }\label{fig:MCRMsig_sum}
\end{figure}

Let $\sigma_{\rm{RM_{sum}}}$ be the sum of the standard deviations for the three inclination angles, $\sigma_{\rm{RM}30}$, $\sigma_{\rm{RM}60}$, and $\sigma_{\rm{RM}90}$, respectively,
\begin{equation}
\label{RMsig_sum}
\sigma_{\rm{RM_{sum}}} =\sqrt{\sigma_{\rm{RM}90}^2+\sigma_{\rm{RM}60}^2+\sigma_{\rm{RM}30}^2},
\end{equation}
we obtain Figure \ref{fig:MCRMsig_sum} and $k \sim 2.9$, which is closer to $k \sim 2.7$ of our Monte-Carlo simulations. This simple test suggests that the deviation from $k\sim 2$ is caused mainly by the quick drop of $\sigma_{\rm{RM90}}$  in redshift. This drop is caused by an enlarge of the physical beam size and a quick decrease of the filling factor for the case of the edge-on. Note that we consider 700 -1800 MHz to determine ${\rm RM_{peak}}$. Even if we consider wider frequency coverage such as 120 -1800 MHz, the result does not significantly change.

Understanding the above statistical property of the DING's contribution to observed RM is crucial to exploring the IGMF. Recent works adopt the idea to see the RM gap between extragalactic polarized sources located at different redshifts \citep{2014PASJ...66...65A} and reduce foreground contributions by comparing random and physical pair-sources \citep{2019ApJ...878...92V, 2020MNRAS.495.2607O}. Here, for random pair-sources, the sources at a higher redshift have more chance to be intervened by DINGs as well as more contribution from the IGMF. It would be difficult to distinguish these contributions from RM data only, remaining a large uncertainty of the estimation of the IGMF. This contamination is also a major problem when we probe the IGMF from the relationship between the RM and redshift of extragalactic sources (\cite{2011ApJ...738..134A}, \cite{2012arXiv1209.1438H}, \cite{2014MNRAS.442.3329X}). Our results suggest that to reduce this uncertainty, we should either focus on polarized sources found at low frequencies and/or select sources that do not show strong depolarization features. In both cases, it is expected no strong DING along the line of sight. Consequently, we obtain the limited number of available sources so far, but future large surveys of RM would deliver the sufficient number of the sources to detect the excess RM caused by the IGMF (see \cite{2014ApJ...790..123A}).

\section{Conclusion}\label{conc}

In order to clarify the contribution of intervening galaxies to the observed polarization properties, we simulated the depolarizing intervening galaxy (DING) along the line of sight when observing a uniform background source. We find that the degree of depolarization depends on the inclination angle and the impact parameter of the DING. We found that the larger the standard deviation, the more likely it is that depolarization will occur. The shape of the Faraday dispersion function (FDF) obtained from RM CLEAN represents the shape of the histogram of RM within the beam and the multi components are clearly visible in the FDF. The appearance of the multi components depends on the model of the DING, but the fraction of the DING that covers the background emission (the filling factor) and the RM structure within the beam are essential. Faraday tomography indicated that the peak Faraday depth of the FDF is different from the beam-averaged RM and the observed RM with the classical style of the DING. We performed Monte-Carlo simulations to understand statistics of polarization properties. The simulations suggest that the standard deviation of observed RMs for the DINGs follows $\sigma_{\rm RM} \propto 1/{(1+z)^k}$ with $k \sim 2.7$.

Finally, the SKA and SKA Precursor/Pathfinder will probe a large number of extragalactic polarized sources that can be depolarized by DINGs discussed in this paper. Although we did not consider observational specifications in detail such as angular resolution and frequency coverage, our simulations indicate that the specifications are very important to accurately evaluate the impact of DINGs on these observations. Such work should be achieved along with the development of survey strategies in the future.

%%%%%%%%%%%%%%%%%%%%%%%%%%%%%%%%%%%%%%%

\begin{ack}
The authors would like to thank X. Sun, S. P. O'Sullivan, B. M. Gaensler, K. Takahashi and H. Sakemi for useful comments for this work. This work was supported in part by JSPS KAKENHI (TA:21H01135,MM:19K03916, 20H01941).
\end{ack}

\appendix
\section{Models}
We consider exponential distribution for the electron density distribution, $n(z)$:
\begin{equation}\label{eq:electron_d}
    n(z) = n_0 \exp{\left( \frac{z}{H} \right)}.
\end{equation}
We adopt $n_0 = 0.014\ \mathrm{cm^{-3}}$ and scale height, $H = 1.83\ \mathrm{kpc}$ \citep{2008PASA...25..184G}.

The magnetic fields (ASS and BSS) can be written in the cylindrical coordinate system as (following the equation in \cite{2008A&A...477..573S}),
\begin{equation}
\left\{
\begin{array}{ll}
B_R(R, \Theta, z) &= D_1(R,z)D_2(R,\Theta)\sin{p} \\
B_{\Theta}(R, \Theta, z) &= - D_1(R,z)D_2(R,\Theta)\cos{p} \\
B_z(R, \Theta, z) &= 0 ,
\end{array}
\right.
\end{equation}
where $p$ is the pitch angle of arms in Milky-way like galaxy, and \begin{equation}
D_1 (R,\Theta) =
\left\{
\begin{array}{ll}
    B_0 \exp{ \left( -\frac{R-R_\odot}{R_0} - \frac{|z|}{z_0} \right) }, & R>R_c, \\
    B_c\exp{ \left(  - \frac{|z|}{z_0} \right) }, &  R \leq R_c.
\end{array}
\right.
\end{equation}

We adopt ASS+RING model for ASS model, and the magnetic field reversals are written in the following way;
\begin{equation}
D_2 (R,\Theta) =
\left\{
\begin{array}{ll}
    +1, & R>7.5\ \mathrm{kpc}, \\
    -1, &  6 < R \leq 7.5\ \mathrm{kpc},\\
    +1, & 5 < R\leq 6\ \mathrm{kpc}, \\
    -1, &  R\leq 5\ \mathrm{kpc},
\end{array}
\right.
\end{equation}
where $+1$ means the clockwise direction as seen from the north pole.
We adopt $R_0 =10\ \mathrm{kpc}$, $z_0 =1\ \mathrm{kpc}$,$R_c =5\ \mathrm{kpc}$, $B_0 =2\ \mathrm{\mu G}$, $B_c =2\ \mathrm{\mu G}$, and $p = -12^{\circ}$.

For BSS model, $D_2$ is written in the following way;
\begin{equation}
    D_2(R, \Theta) = \sin{\left(\Theta + \frac{1}{\tan{p}} + \ln{\frac{R}{R_{\rm sb}}}  \right)}.
\end{equation}
We adopt $R_0 = 6\ \mathrm{kpc}$, $z_0 =1\ \mathrm{kpc}$,$R_c =3\ \mathrm{kpc}$, $B_0 = 2\ \mathrm{\mu G}$ and $B_c =2\ \mathrm{\mu G}$. We also set $R_{\rm sb} = 9\ \mathrm{kpc}$ and $p = -10^{\circ}$ for $R > 6\ \mathrm{kpc}$, and otherwise $R_{\rm sb} = 6\ \mathrm{kpc}$ and $p = -15^{\circ}$.

The all-aky RM maps suggest that milky-way galaxy has the the existence of halo toroidal fields because we observe reverse RM structure across the galactic plane and across the galactic center (e.g., \cite{2009ApJ...702.1230T}). The halo toroidal field is written in the following way \citep{2003A&A...410....1P}:
\begin{equation}
\left\{
\begin{array}{ll}
B_{tR}(R, \Theta, z) &= 0 \\
B_{t\Theta}(R, \Theta, z) &= B_{t0}\frac{{\rm sign}(z)^v}{1+\left(\frac{|z|-z_{t0}}{z_{t1}} \right)^2}\frac{R}{R_{t0}}\exp{\left(-\frac{R - R_{t0}}{R_{t0}} \right)} \\
B_{tz}(R, \Theta, z) &= 0 ,
\end{array}
\right.
\end{equation}
where ${\rm sign}(z)$ is the sign of $z$, the $v$ is the parity of the toroidal field configuration. If we consider the asymmetric configuration in longitude and latitude relative to the Galactic plane and centre and the axisymmetric configuration without reversals relative to the Galactic plane, we adopt $v = 1$ and $v = 2$, respectively.The parameters are $B_{t0} = 2 \mathrm{\mu G}$, $R_{t0} = 4 \mathrm{kpc}$, $z_{t0} = 1.5 \mathrm{kpc}$, and $z_{t1} = 0.2 \mathrm{kpc}$ for $|z| < z_{t0}$ and otherwise $z_{t1} = 4 \mathrm{kpc}$. The paranerters are based on \citet{2010RAA....10.1287S}.

%%%
% See the manual for the detail.
%%%
\DeclareAbbreviation\nar{New Astronomy Review}
\DeclareAbbreviation\pasa{PASA}

\end{document}